\newcommand{\kms}{{km$~\!$s$^{-1}$}}
\newcommand{\ang}{{\AA}}

\newcommand{\msol}{\mathcal{M}_\odot}

\newcommand{\lksol}{{L}_\odot^K}
\newcommand{\sdu}{$\msol~\!$pc$^{-2}$}

\newcommand{\muu}{mag arcsec$^{-2}$}

\newcommand{\hone}{H{\scshape$~\!$i}}

\newcommand{\molh}{H$_2$}

\newcommand{\pV}{\mathcal{V}}

\newcommand{\sigr}{\sigma_R}

\newcommand{\xco}{X_{\rm CO}}

\newcommand{\sdhi}{\Sigma_{\mbox{\rm \footnotesize H{\scshape i}}}}

\newcommand{\rhi}{R_{\mbox{\rm \footnotesize H{\scshape i}}}}

\newcommand{\sdmh}{\Sigma_{\rm H_2}}

\newcommand{\mhi}{\mathcal{M}_{\mbox{\rm \footnotesize H{\scshape i}}}}
\newcommand{\mht}{\mathcal{M}_{\mbox{\footnotesize \molh}}}

\newcommand{\qgas}{Q_g}
\newcommand{\qstar}{Q_\ast}
\newcommand{\qrw}{Q_{\rm RW}}

\newcommand{\sfr}{\dot{\mathcal{M}}_\ast}

\newcommand{\Ssfr}{\dot{\Sigma}_\ast}
\newcommand{\Ssfre}{\dot{\Sigma}_{e,\ast}}
\newcommand{\Sge}{\Sigma_{e,g}}
\newcommand{\Sse}{\Sigma_{e,\ast}}

\newcommand{\qrwmin}{\qrw^{\rm min}}
\newcommand{\bth}{\boldsymbol\theta}
\newcommand{\bph}{\boldsymbol\phi}
\newcommand{\bvp}{\boldsymbol\varphi}
\newcommand{\bps}{\boldsymbol\psi}

\documentclass[apj,twocolappendix,numberedappendix]{emulateapj}
\RequirePackage{calc}
\usepackage{natbib}
\usepackage{url}
\usepackage{amsmath}

\slugcomment{Accepted: 4 Feb 2014}

\shortauthors{Westfall et al.}
\shorttitle{The DiskMass Survey. VIII.}

\defcitealias{2010ApJ...716..198B}{Paper~I}
\defcitealias{2010ApJ...716..234B}{Paper~II}
\defcitealias{2011ApJS..193...21W}{Paper~III}
\defcitealias{2011ApJ...742...18W}{Paper~IV}
\defcitealias{2011ApJ...739L..47B}{Paper~V}
\defcitealias{2013A&A...557A.130M}{Paper~VI}
\defcitealias{2013A&A...557A.131M}{Paper~VII}
\defcitealias{1998ApJ...498..541K}{K98}
\defcitealias{2010ApJ...721..975O}{OML10}

\begin{document}

\title{ The DiskMass Survey. VIII. \\ On the Relationship Between Disk Stability
and Star Formation }

\author{ Kyle B. Westfall\altaffilmark{1,2}, David R. Andersen\altaffilmark{3},
Matthew A. Bershady\altaffilmark{4},\\ Thomas P. K. Martinsson\altaffilmark{5},
Robert A. Swaters\altaffilmark{6}, \& Marc A. W. Verheijen\altaffilmark{1} }

\altaffiltext{1}{Kapteyn Astronomical Institute, University of Groningen,
Landleven 12, 9747 AD Groningen, the Netherlands}

\altaffiltext{2}{National Science Foundation (USA) International Research
Fellow}

\altaffiltext{3}{NRC Herzberg Institute of Astrophysics, 5071 West Saanich Road,
Victoria, BC V9E 2E7, Canada}

\altaffiltext{4}{Department of Astronomy, University of Wisconsin-Madison, 475
N. Charter St., Madison, WI 53706, USA}

\altaffiltext{5}{Leiden Observatory, Leiden University, PO Box 9513, 2300 RA
Leiden, The Netherlands}

\altaffiltext{6}{National Optical Astronomy Observatory, 950 North Cherry Ave,
Tucson, AZ 85719, USA }

\email{westfall@astro.rug.nl}

\begin{abstract}

We study the relationship between the stability level of late-type galaxy disks
and their star-formation activity using integral-field gaseous and stellar
kinematic data.  Specifically, we compare the two-component (gas$+$stars)
stability parameter from \citeauthor{2011MNRAS.416.1191R} ($\qrw$),
incorporating stellar kinematic data for the first time, and the star-formation
rate estimated from 21cm continuum emission.  We determine the stability level
of each disk probabilistically using a Bayesian analysis of our data and a
simple dynamical model.  Our method incorporates the shape of the stellar
velocity ellipsoid (SVE) and yields robust SVE measurements for over 90\% of our
sample.  Averaging over this subsample, we find a meridional shape of
$\sigma_z/\sigma_R = 0.51^{+0.36}_{-0.25}$ for the SVE and, at 1.5 disk scale
lengths, a stability parameter of $\qrw = 2.0\pm 0.9$.  We also find that the
disk-averaged star-formation-rate surface density ($\Ssfre$) is correlated with
the disk-averaged gas and stellar mass surface densities ($\Sge$ and $\Sse$) and
anti-correlated with $\qrw$.  We show that an anti-correlation between $\Ssfre$
and $\qrw$ can be predicted using empirical scaling relations, such that this
outcome is consistent with well-established statistical properties of
star-forming galaxies.  Interestingly, $\Ssfre$ is not correlated with the
gas-only or star-only \citeauthor{1964ApJ...139.1217T} parameters, demonstrating
the merit of calculating a multi-component stability parameter when comparing to
star-formation activity.  Finally, our results are consistent with the
\citeauthor{2010ApJ...721..975O} model of self-regulated star-formation, which
predicts $\Ssfre/\Sge\propto\Sse^{1/2}$.  Based on this and other theoretical
expectations, we discuss the possibility of a physical link between disk
stability level and star-formation rate in light of our empirical results.

\end{abstract}


\section{Introduction}
\label{sec:intro}

Stars are formed by the collapse of gas.  In galaxy disks, a gas cloud should be
gravitationally unstable if (1) it cannot adjust its internal pressure to
balance the local gravitational pressure on timescales shorter than a free-fall
time and (2) it occupies an area smaller than the scale on which differential
rotation will shear it apart.  \citet{1964ApJ...139.1217T} codified these
concepts into a criterion for the stability of an infinitely thin, rotating,
self-gravitating, fluid disk:
\begin{equation}
Q = \frac{\kappa \sigma}{\pi G \Sigma} > 1,
\label{eq:q}
\end{equation}
where
\begin{equation}
\kappa^2 = 2\frac{v_c}{R}\left( \frac{v_c}{R} + \frac{\partial v_c}{\partial
R}\right)
\label{eq:efreq}
\end{equation}
is the epicyclic frequency, $v_c$ is the circular speed of the potential,
$\sigma$ is the radial velocity dispersion, $\Sigma$ is the mass surface
density, and $G$ is the gravitational constant.  However, self-gravity is only
one major player in the star-formation process, with chemodynamical processes,
turbulence, and magnetic fields also significantly affecting the dynamics
\citep[see][and references therein]{2007ARA&A..45..565M}.  The relative
importance of these physical properties to the star-formation law is a matter of
ongoing debate.

Empirical studies of the star-formation law in disk galaxies have predominantly
focused on observations of their gaseous components.  In a seminal article,
\citet[][hereafter \citetalias{1998ApJ...498..541K}]{1998ApJ...498..541K}
demonstrated that the star-formation rate per unit area ($\Sigma_{\rm SFR}$), or
equivalently the time derivative of the stellar mass surface density ($\Ssfr
\equiv \Sigma_{\rm SFR}$), is well correlated with the surface density of the
hydrogen gas, $\Sigma_{\rm H} = \sdhi+\sdmh$, following the star-formation law
suggested by \citet{1959ApJ...129..243S}.  Considering the proportionality from
equation \ref{eq:q}, one might expect such a relation if star-formation is
driven by self-gravity.  Owing much to the flood of relevant data, the
quantitative details of the star-formation law in galaxy disks and its relation
to the Kennicutt-Schmidt (KS) law \citepalias[$\Ssfr \propto \Sigma_{\rm
H}^{1.4}$;][]{1998ApJ...498..541K} have been greatly scrutinized.  This scrutiny
has lead to a number of alternatives to this paradigm; see compilations by,
e.g., \citet{2008AJ....136.2782L} and \citet{2013MNRAS.434.3389Z}.

Many of these alternatives involve consideration of gas properties, such as
accounting for the dust-to-gas ratio \citep{2013AJ....146...19L}, or use of
specific gas tracers.  \citet{2002ApJ...569..157W} and
\citet{2008AJ....136.2846B} have shown that $\Ssfr$ is more tightly correlated
with $\sdmh$ than with $\Sigma_{\rm H}$.  Therefore, the star-formation law will
also be affected by the ability of the interstellar medium to convert \hone\ to
H$_2$ \citep{2004ApJ...612L..29B, 2006ApJ...650..933B}, which can be related to
the hydrostatic pressure in the disk plane \citep{1989ApJ...338..178E,
1993ApJ...411..170E}.  As an inherently dynamical process, the star-formation
law should also be influenced by relevant dynamical timescales, such as the
local free-fall time \citep{2012ApJ...745...69K}.
\citetalias{1998ApJ...498..541K} considered a star-formation law that
incorporated the orbital timescale \citep{1997ApJ...481..703S,
1997RMxAC...6..165E}, which is relevant in a scenario where dynamical processes
in the disk (such as bars and spiral arms) are a primary driver of star
formation.

Here, we explore the role of the disk stability level in the star-formation
process.  Although the star-formation law may be written to show an explicit
dependence on the {\it gas} stability parameter \citep[$Q_g$; see,
e.g.,][]{2005ApJ...630..250K}, the most relevant assessment of the stability
level includes both the gas and the stars \citep{2005ApJ...626..823L}.  Indeed,
\citet{2003MNRAS.346.1215B} have shown that the two-component (gas$+$stars)
stability parameter of 16 galaxies is a better estimator of the star-formation
{\em threshold} than one incorporating the gas alone.  In general, however, the
effect of the stellar component on the star-formation {\em law} is not
well-understood:  \citet{2003MNRAS.346.1215B} have also shown that a
star-formation law that considers only the gas component (the KS law) is
statistically indistinguishable from one proposed by \citet{1994ApJ...430..163D}
that incorporates the {\em total} disk mass surface density.  More recently,
\citet{2011ApJ...733...87S} have shown an explicit dependence of $\Ssfr$ on
$\Sigma_\ast$.  Their ``extended Schmidt law'' --- one that incorporates the
dependence on $\Sigma_\ast$ --- is consistent with the self-regulation model
proposed by \citet[][hereafter
\citetalias{2010ApJ...721..975O}]{2010ApJ...721..975O}, who find $\Ssfr \propto
\Sigma_g \Sigma_\ast^{1/2}$ for galaxies with a constant-scale-height stellar
disk\footnote{
In detail, the star-formation law from \citetalias{2010ApJ...721..975O} allows
for star formation in starless systems, which would be prohibited by a law with
an explicit dependence on the stellar mass of a galaxy.
} \citep[see also][]{2011ApJ...743...25K, 2013ApJ...776....1K}.  In their model,
the explicit dependence of the star-formation law on $\Sigma_\ast$ is via its
contribution to the vertical gravitational field of the disk.  Therefore, we
also consider the correlation between the star-formation activity of a disk and
its stellar mass surface density.

Previous studies considering the relation of the two-component stability level
of disks and/or stellar mass surface density to the star-formation law have
lacked the kinematic data necessary to measure either of these quantities
dynamically.  Instead, they have used stellar mass estimates from
stellar-population-synthesis modeling, which have not been directly calibrated
by dynamical mass measurements in external disk galaxies \citep[see discussion
in][hereafter \citetalias{2010ApJ...716..198B}]{2010ApJ...716..198B}.  However,
with its unparalleled stellar kinematic data in the dynamically cold regime of
galaxy disks and its ancillary gas data, the DiskMass Survey
\citepalias{2010ApJ...716..198B} is well suited to studying the effect of the
stellar component (via its mass surface density and stability level) on star
formation in galaxy disks.

Our paper is organized as follows:  We briefly discuss the relevant
observational data in Section \ref{sec:data}.  We describe our dynamical
modeling in Section \ref{sec:genmodel}; however, a more detailed discussion of
this modeling approach will be presented in a forthcoming paper.  For now, we
provide a brief summary of the equations used in the dynamical model in Appendix
\ref{app:dynmodel}, and we discuss the details of our sampling of the posterior
probability of the model in Appendix \ref{app:converge}.  Our probabilistic
modeling is the basis for our calculations of the disk stability parameter and
stellar mass surface density.  These calculations, the stability results, and a
comparison of the star-formation properties of our galaxy sample with the
``Normal Spirals'' from \citetalias{1998ApJ...498..541K} are discussed in
Section \ref{sec:stability}.  We explore any correlations among disk stability
level, stellar mass surface density, and star-formation rate in Section
\ref{sec:sfrcorr}.  Among other findings, we show that the two-component disk
stability parameter is anti-correlated with the star-formation activity of the
disk.  In Section \ref{sec:scaling}, we show that this anti-correlation can be
predicted by considering a closed system of empirical scaling relations.
Finally, we summarize and briefly discuss our results in Section
\ref{sec:conclusions}.

\section{ Observational Data }
\label{sec:data}

Our galaxy sample is described by \citet[][hereafter
\citetalias{2013A&A...557A.130M}]{2013A&A...557A.130M}.  However, here we limit
our analysis to the 27 galaxies with available measurements of the 21cm
radio-continuum flux density, $S_{21}$, which we use as our star-formation-rate
estimator (see Section \ref{sec:qeqns}).  Fifteen measurements of $S_{21}$ are
taken from our Survey data \citep{TPKMPhD} and the remaining twelve are drawn
from the NRAO/VLA Sky Survey \citep[NVSS;][]{1998AJ....115.1693C};\footnote{
\url{http://www.cv.nrao.edu/nvss/}
} when data were available from both NVSS and the DiskMass Survey (nine
galaxies), we chose the measurement with the smallest error.

The additional data products used in this study are (1) SparsePak\footnote{
Mounted on the 3.5-meter WIYN telescope, a joint facility of the University of
Wisconsin-Madison, Indiana University, Yale University, and the National Optical
Astronomy Observatories.
} integral-field spectroscopy (IFS) from 6480-6890 \ang\ (at a resolution of
$\lambda/\delta\lambda \sim 11500$) used to obtain ionized-gas kinematics, (2)
PPak\footnote{
Mounted with PMAS on the 3.5-meter telescope at the Calar Alto Observatory,
operated jointly by the Max-Planck-Institut für Astronomie (MPIA) in Heidelberg,
Germany, and the Instituto de Astrofísica de Andalucía (CSIC) in Granada, Spain.
} IFS from 4975-5375 \ang\ ($\lambda/\delta\lambda \sim 7700$) used to obtain
stellar kinematics \citepalias{2013A&A...557A.130M}, (3) {\it Spitzer} MIPS
imaging at 24$\mu$m used to obtain the molecular-mass surface density
\citep[][hereafter \citetalias{2011ApJ...742...18W}]{2011ApJ...742...18W}, and
(4) Westerbork and Very Large Array (VLA) radio synthesis imaging of the 21cm
emission line used to obtain the atomic-mass surface density.  Atlases of the
data are provided by \citet{TPKMPhD}, \citetalias{2013A&A...557A.130M}, and
\citet[][hereafter \citetalias{2013A&A...557A.131M}]{2013A&A...557A.131M}.  For
three of the galaxies, we have not yet determined the atomic-mass surface
densities observationally.  For these galaxies, we approximate the atomic-mass
surface density following the procedure provided in Section 3.1 of
\citetalias{2013A&A...557A.131M}.

We refer the reader to the referenced papers for a full description of our
handling of the raw data and the subsequent analysis leading to our primary data
products.  However, we make two brief comments: (1) Although slightly modified,
the determination of the 24$\mu$m surface brightness profiles and the subsequent
calculation of the molecular-mass surface density is nearly identical to the
analyses done in Papers IV and VII.  (2) Using $^{12}$CO(1--0) observations of
five galaxies in our sample obtained from 7--9 Jan 2012 using the IRAM 30m
telescope,\footnote{
IRAM (Institut de Radioastronomie Millim\'{e}trique) is supported by INSU/CNRS
(France), MPG (Germany) and IGN (Spain).
} our preliminary analysis confirms that the 24$\mu$m-to-CO calibration from
\citetalias{2011ApJ...742...18W} has an error of roughly 30\%, which is included
in our error analysis.  This error is systematic for individual galaxies, but
the systematic errors are distributed normally within our full sample.
Throughout this paper we assume $H_0 = 73\pm5$ \kms\ Mpc$^{-1}$, $\Omega_M =
0.27$, $\Omega_\Lambda = 0.73$, and we adopt a CO-H$_2$ conversion factor of
$\xco = (2.7 \pm 0.9) \times 10^{20}$ cm$^{-2}$ (K \kms)$^{-1}$
\citepalias{2011ApJ...742...18W}.

\section{ Analysis }
\label{sec:stability}

Our results hinge on the dynamical modeling of our integral-field data, which
follows a holistic, Bayesian approach.  A full description of this approach is
beyond our present scope and will be presented in a forthcoming paper.  Here,
we discuss the basic setup of our dynamical model in Section \ref{sec:genmodel},
along with a brief outline of the analytic equations in Appendix
\ref{app:dynmodel} and a detailed description of how we produce samples of the
probabilistic model in Appendix \ref{app:converge}. The goal of our modeling for
this paper is to constrain the disk stability level and dynamical mass surface
density, $\Sigma_{\rm dyn}$, as a function of radius for each galaxy.  However,
these quantities are not explicit elements of our dynamical model.  Instead,
they are calculated using the posterior distribution of the model, as described
in Section \ref{sec:qeqns}.  Section \ref{sec:qeqns} also discusses our
star-formation-rate measurements.  As a reference point for the subsequent
discussion of the star-formation activity, we compare our galaxy sample to the
set of ``Normal Spirals'' from \citetalias{1998ApJ...498..541K} and the
expectation of the KS law in Section \ref{sec:KS}.  Finally, we discuss our disk
stability results in Section \ref{sec:result}.

\subsection{ Probabilistic Modeling }
\label{sec:genmodel}

Our dynamical assumptions are virtually identical to those from
\citetalias{2011ApJ...742...18W}; however, our analysis is now done following
Bayesian statistics.  The statistical background provided by \citet{MacKay:itp},
the practical examples provided by \citet{2010arXiv1008.4686H}, and the
sampling algorithm provided by \citet[][and extensions
thereof]{2013PASP..125..306F} have been invaluable resources in our application
of this approach.

In Appendix \ref{app:dynmodel}, we briefly present the defining equations of our
dynamical model, derived by adopting a set of hypotheses, $\mathcal{H}$, that
result in a set of parameters, $\bth$.  The goal of our fitting procedure is to
determine the probability, $P$, that a model with parameters $\bth$ could have
generated our observational data, $\mathcal{D}$.  That is, our goal is to obtain
the conditional probability $P(\bth|\mathcal{D},\mathcal{H})$, read as ``the
probability of $\bth$ given $\mathcal{D}$ and $\mathcal{H}$'' and termed the
posterior probability.  We calculate the posterior probability using Bayes'
theorem,
\begin{equation}
P({\bth} | \mathcal{D}, \mathcal{H}) \propto P(\mathcal{D} | {\bth},
\mathcal{H}) P({\bth} | \mathcal{H}),
\label{eq:bayes}
\end{equation}
where $\mathcal{L} = P(\mathcal{D} | {\bth}, \mathcal{H})$ is the likelihood of
the model and $P({\bth} | \mathcal{H})$ is the prior probability of the model.
Our analysis ignores the proportionality constant, $P(\mathcal{D} |
\mathcal{H})$, called the ``evidence'' or ``marginal likelihood,''  which is the
integral of the right-hand side of equation \ref{eq:bayes} over the full
parameter space.  We ignore the ``evidence'' because its primary use is in
comparing hypotheses (``model comparison''), which we have not done for this
paper.

Calculations using equation \ref{eq:bayes} are analytic for our generative ---
fully probabilistic --- model.  The dynamical model described in Appendix
\ref{app:dynmodel} produces all the line-of-sight (LOS) kinematics --- stellar
velocity, $V_\ast$; stellar velocity dispersion, $\sigma_\ast$; ionized-gas
velocity, $V_g$; and ionized-gas velocity dispersion, $\sigma_{ig}$ --- and the
radial profile of the cold-gas mass surface density, $\Sigma_g = 1.4 (\sdhi +
\sdmh)$.  These model quantities are compared to the data using the likelihood
function, $\mathcal{L}$, which we define as the product of all Gaussian
probabilities representing the data.  Our generative model includes intrinsic
scatter in $V_\ast$, $\sigma_\ast$, $V_g$, and $\sigma_{ig}$, but not
$\Sigma_g$.  The uncertainty in $\Sigma_g$ is dominated by systematic error such
that intrinsic scatter is contraindicated.  The variance of each kinematic
measurement used to calculate $\mathcal{L}$ is thus the quadrature sum of the
measurement error and the relevant intrinsic scatter \citep[see, e.g., equations
9, 10, and 35 from][]{2010arXiv1008.4686H}.  Our inclusion of intrinsic scatter
ensures that the posterior probability is not strongly affected by stochastic
deviations of the data about our simplistic model.

For most model parameters, we adopt (nominally) ``noninformative'' priors
\citep[either linearly or log-linearly uniform; see][]{MacKay:itp} with upper
and lower limits that have effectively zero posterior probability.  To constrain
the inclination, however, we assume our galaxies follow the Tully-Fisher (TF)
relation from \citet{2001ApJ...563..694V}.  The inclination is therefore not an
explicit parameter, but calculated using equation 3 from
\citetalias{2011ApJ...742...18W}.  The absolute $K$-band magnitude and TF
zero-point used in this calculation are normally distributed about their
measured value according to the measurement error (see Table 5 from
\citetalias{2013A&A...557A.130M}) and TF scatter (0.27 dex), respectively.  Our
homoscedastic TF scatter is based on a conservative estimate of the intrinsic
scatter in the relation and the distance error for the Ursa Major cluster.  This
TF prior is critical to the projection calculations for galaxies with
inclinations lower than $\sim$20 degrees \citep[cf.][]{2013ApJ...768...41A}.
Although it is rarely an issue, we also force $\Sigma_{\rm dyn} > \Sigma_g$ at
1.5 scale lengths; we assume $\Sigma_{\rm dyn} = \Sigma_g + \Sigma_\ast$ such
that this constraint forces $\Sigma_\ast > 0$.

Although the calculations of the posterior probability based on $\mathcal{L}$
and our chosen priors are analytic, the statistics relevant to our discussion
below, such as the median and confidence intervals of the posterior probability
marginalized over specific parameters, require integrals of equation
\ref{eq:bayes} that are non-trivial.  Therefore, we use a Markov Chain Monte
Carlo (MCMC) method to generate coordinates $\bth$ that are drawn in
proportion to the posterior probability.  With such samples, it becomes
straight-forward to perform the relevant integrals by computing cumulative
distributions in one or more dimensions.

To sample from the posterior, we use the stretch-move MCMC sampler from
\citet{2013PASP..125..306F} in combination with a parallel-tempering algorithm
as implemented by these authors (see \url{http://dan.iel.fm/emcee/}).  The
parallel-tempering scheme proves to better sample probability densities that
exhibit significant curvature --- non-linear correlations between parameters in
the model.  Our analysis uses our own {\tt C++} implementation of these
algorithms.  We provide the detailed method we use to produce the samples of the
posterior probability in Appendix \ref{app:converge}.

Our prior for the meridional shape of the SVE is uniform in the range
$0.01\leq\alpha=\sigma_z/\sigma_R\leq3.0$; however, we do not expect $\alpha$ to
be greater than one.  Throughout the remainder of this paper, we omit two
galaxies (out of 27) from consideration --- UGC 7917 and 11318 --- because our
probabilistic modeling has produced unsatisfactory constraints on $\alpha$.  UGC
11318 is very nearly face-on ($i\sim6\arcdeg$) such that the in-plane motions
are highly projected, likely leading to an erroneous $\alpha$ that is
significantly larger than unity ($\alpha=1.9^{+0.5}_{-0.4}$).  For UGC 7917, the
marginalized probability distribution for $\alpha$ is biased by our prior
assumption of $\alpha < 3$, suggesting that the likelihood function would prefer
values that are even larger than this limit.  The reason for this unphysical
result is unclear; however, we note that UGC 7917 has a strong bar and is
largely devoid of ionized gas near its center.  This yields a poor measurement
of asymmetric drift, which is a crucial measurement in our dynamical model.  We
find reasonable assessments of the disk SVE for the remaining subsample of 25
galaxies; these galaxies show a marginalized median and 68\% confidence interval
of $\alpha = 0.51^{+0.36}_{-0.25}$.  A full discussion of our SVE results will
be the focus of a forthcoming paper.

Finally, we note that the probabilistic modeling discussed so far only produces
posterior distributions for each parameter in our dynamical model.  However,
neither all quantities of interest nor all elements of the calculations
discussed in the next section are direct parameters of that model.  For example,
we calculate the scale height, $h_z$, based on the measured scale length, $h_R$,
using the oblateness relation from \citet[][hereafter
\citetalias{2010ApJ...716..234B}; see their equation 1]{2010ApJ...716..234B}; in
this example, we adopt a normal distribution for $h_R$ according to its measured
value and error.  There are three classes of derived quantities of interest: (1)
$\bph$ --- those that are independent of any other quantity (e.g., $h_R$); (2)
$\bvp$ --- those that are only dependent on $\bph$ (e.g., $h_z$); and (3) $\bps$
--- those that are dependent on both the new quantities and the parameters of
our dynamical model (e.g., $\Sigma_{\rm dyn}$).  Given that $\mathcal{L}$ is
only dependent on $\bth$, the posterior probability that includes these extra
quantities is
\begin{equation}
P(\bph, \bvp, \bps, \bth | \mathcal{D}) = P(\bth | \mathcal{D})\ P(\bps | \bph,
\bvp, \bth)\ P(\bvp | \bph)\ P(\bph),
\label{eq:pfull}
\end{equation}
where we have omitted the dependence on $\mathcal{H}$ for clarity.  Therefore,
we determine the posterior probability of each derived quantity using the
samples of the posterior probability of our dynamical model and samples of the
additional {\em known} prior probability distributions, $P(\bph)$.  Unless
otherwise noted, such as the $\Sigma_{\rm dyn} > \Sigma_g$ constraint discussed
above, the prior probabilities $P(\bps | \bph, \bvp, \bth, \mathcal{H})$ and
$P(\bvp | \bph, \mathcal{H})$ in equation \ref{eq:pfull} are assumed to be
uniform.

\subsection{ Calculations of Disk Stability, Mass Surface Density, and Star-Formation Rate }
\label{sec:qeqns}

We assume each galaxy disk consists of two components, a thin cold-gas disk and
a thin stellar disk.  We assume the mass surface densities of all other disk
components (such as the thick stellar disk) are much smaller with stability
levels that are much higher than either of these two components.  The atomic-
and molecular-gas disks are subsumed into a single cold-gas disk, with the
implied assumption being that they have roughly the same vertical mass
distribution and velocity dispersion \citep{2013AJ....146..150C}.  

The stability parameter we calculate here, generally signified by $Q$ as in
equation \ref{eq:q}, was derived by \citet[][see also
\citealt{2013MNRAS.433.1389R}]{2011MNRAS.416.1191R}.  In their formulation, one
corrects the stability parameter for the disk thickness by calculating
\begin{equation}
\mathcal{T}_j = \left\{
\begin{array}{rl}
0.8 + 0.7\ \alpha_j & {\rm for}\ \ 0.5 < \alpha_j \leq 1 \\[6pt]
1.0 + 0.6\ \alpha_j^2 & {\rm for}\ \ 0.0 < \alpha_j \leq 0.5
\end{array}
\right.
\label{eq:thickq}
\end{equation}
such that $\mathcal{T}_j\geq1$ for each component $j$, where $\alpha_j =
\sigma_{z,j}/\sigma_{R,j}$ is the meridional shape of its velocity ellipsoid.
For the gas, we assume the velocity ellipsoid is isotropic such that
$\mathcal{T}_g = 1.5$.  The thickness-corrected stability parameter is
$\mathcal{T}_j Q_j$; from equation \ref{eq:q}, $Q_g = \kappa \sigma_{cg}/(\pi G
\Sigma_g)$, where $\sigma_{cg}$ is the velocity dispersion of the {\em cold}
gas, and $Q_\ast = \kappa \sigma_R/(\pi G \Sigma_\ast)$.  The thickness
corrections do not strongly depend on the assumed vertical density profile or
the oblateness, within the empirical expectations for these disk properties
\citep[][private communication]{1992MNRAS.256..307R}.  The two-component
(gas$+$stars) stability parameter is then
\begin{equation}
\qrw^{-1} = \left\{
\begin{array}{ll}
w_{\sigma}/\mathcal{T}_\ast \qstar + 1/\mathcal{T}_g \qgas & {\rm for}\
\mathcal{T}_\ast \qstar > \mathcal{T}_g \qgas \\[6pt]
1/\mathcal{T}_\ast \qstar + w_{\sigma}/\mathcal{T}_g \qgas & {\rm for}\
\mathcal{T}_\ast \qstar < \mathcal{T}_g \qgas
\end{array}
\right.,
\label{eq:qrw}
\end{equation}
where the weight of the most stable component is
\begin{equation}
w_{\sigma} = \frac{2 \sigr \sigma_{cg}}{\sigr^2 + \sigma^2_{cg}}.
\label{eq:qweight}
\end{equation}
%
Thus, calculations of $\qrw$ require the circular-speed curve ($v_c$ is needed
to obtain the epicyclic frequency, $\kappa$), the radial velocity dispersion of
the cold gas and stars ($\sigma_{cg}$ and $\sigma_R$), and the mass surface
densities of the cold gas and stars ($\Sigma_g$ and $\Sigma_\ast$); $v_c$ and
$\sigma_R$ are direct products of the dynamical model.

Our data only provide the ionized-gas dispersion, $\sigma_{ig}$, whereas our
disk-stability calculations incorporate the cold-gas velocity dispersion,
$\sigma_{cg}$.  We can roughly match the mode of the distribution of H$\alpha$
velocity dispersions from \citet[][Figure 6]{2006ApJS..166..505A} to that of the
\hone\ and CO velocity dispersion distribution from \citet[][Figure
5]{2013AJ....146..150C} by setting $\sigma_{cg} \approx \sigma_{ig}/2$.  This
assumption yields cold-gas velocity dispersions of $6.5\leq \sigma_{cg} \leq
10.5$ \kms\ for our sample, which is consistent with direct measurements
\citep[e.g.,][]{2012AJ....144...96I}.  As evident from equation \ref{eq:q},
systematic errors in this assumption yield equivalent systematic errors in
$Q_g$.

We obtain direct measurements of $\Sigma_g$ from our observations
\citepalias[see][]{2013A&A...557A.131M}; however, we use our model
parameterization of $\Sigma_g$ in the following analysis for consistency (e.g.,
between $\Sigma_g$ and $v_c$, see Appendix \ref{app:dynmodel}).  The stellar
surface mass density is $\Sigma_\ast = \Sigma_{\rm dyn} - \Sigma_g$, where we
use $\Sigma_{\rm dyn} = \sigma_z^2 / (\pi k G h_z)$ to calculate the dynamical
mass surface density \citep[see][and earlier papers in this
series]{1988A&A...192..117V}.  The integration constant $k$ depends on the
assumed vertical mass density profile.  Following the discussion in Section
2.2.1 of \citetalias{2010ApJ...716..234B}, we adopt $k=1.5$, the integration
constant for the purely exponential vertical mass distribution $\rho(z) \propto
\exp(-z/h_z)$.  We assume the scale height, $h_z$, is constant at all radii and
we calculate its value using the measured $h_R$ and the oblateness relation from
\citetalias{2010ApJ...716..234B}.  In our probabilistic model, we adopt normal
distributions for the scale length and the zero-point of the oblateness
relation.

\begin{deluxetable*}{ r r r r r r r r r }
\tabletypesize{\scriptsize}
\tablewidth{0pt}
\tablecaption{ Effective Star-formation Rates, Mass Surface Densities, and Disk Stabilities }
\tablehead{
 & \colhead{Hubble} & \colhead{$D_{25}$\tablenotemark{b}} & \colhead{$\log\Ssfre$} & \colhead{$\log\Sge$} & \colhead{$\log\Sse$} & \colhead{$\log(\Ssfre/\Sge)$} & \colhead{$\log(\Ssfre\Sge^{-1}\Sse^{-1/2})$} & \\
 \colhead{UGC} & \colhead{Type\tablenotemark{a}} & \colhead{[arcsec]} & \colhead{[$\msol$ pc$^{-2}$ Gyr$^{-1}$]} & \colhead{[$\msol$ pc$^{-2}$]} & \colhead{[$\msol$ pc$^{-2}$]} & \colhead{[Gyr$^{-1}$]} & \colhead{[$(G/{\rm pc})^{1/2}$]} & \colhead{$\log\qrwmin$} }
\startdata
   448 &  SABc &  99.6 & $   0.11^{+0.15}_{-0.19} $ & $   0.93^{+0.02}_{-0.02} $ & $   1.52^{+0.14}_{-0.15} $ & $  -0.83^{+0.15}_{-0.19} $ & $  -3.42^{+0.17}_{-0.20} $ & $   0.25^{+0.06}_{-0.08} $ \\
   463 &  SABc &  99.6 & $   1.10^{+0.12}_{-0.16} $ & $   1.36^{+0.04}_{-0.04} $ & $   1.64^{+0.15}_{-0.18} $ & $  -0.25^{+0.12}_{-0.16} $ & $  -2.90^{+0.15}_{-0.18} $ & $   0.13^{+0.06}_{-0.06} $ \\
  1087 &    Sc &  90.8 & $   0.47^{+0.14}_{-0.18} $ & $   0.82^{+0.02}_{-0.02} $ & $   1.02^{+0.18}_{-0.22} $ & $  -0.35^{+0.15}_{-0.18} $ & $  -2.68^{+0.18}_{-0.20} $ & $   0.44^{+0.07}_{-0.07} $ \\
  1529 &    Sc & 106.7 & $  -0.16^{+0.15}_{-0.18} $ & $   0.92^{+0.02}_{-0.02} $ & $   1.15^{+0.18}_{-0.22} $ & $  -1.08^{+0.15}_{-0.18} $ & $  -3.48^{+0.18}_{-0.20} $ & $   0.40^{+0.05}_{-0.05} $ \\
  1635 &   Sbc & 111.7 & $  -0.15^{+0.17}_{-0.22} $ & $   0.58^{+0.02}_{-0.02} $ & $   1.01^{+0.18}_{-0.19} $ & $  -0.73^{+0.17}_{-0.22} $ & $  -3.06^{+0.20}_{-0.24} $ & $   0.50^{+0.10}_{-0.11} $ \\
  1908 &   SBc &  79.1 & $   0.95^{+0.12}_{-0.16} $ & $   1.21^{+0.04}_{-0.04} $ & $   1.58^{+0.22}_{-0.34} $ & $  -0.27^{+0.13}_{-0.16} $ & $  -2.88^{+0.21}_{-0.21} $ & $   0.30^{+0.06}_{-0.06} $ \\
  3140 &    Sc & 117.0 & $   0.57^{+0.12}_{-0.16} $ & $   1.11^{+0.02}_{-0.03} $ & $   2.07^{+0.17}_{-0.15} $ & $  -0.54^{+0.13}_{-0.16} $ & $  -3.40^{+0.15}_{-0.18} $ & $   0.02^{+0.10}_{-0.10} $ \\
  3701 &   Scd & 114.3 & $  -0.47^{+0.16}_{-0.20} $ & $   0.85^{+0.01}_{-0.01} $ & $   0.51^{+0.48}_{-0.80} $ & $  -1.32^{+0.16}_{-0.20} $ & $  -3.47^{+0.34}_{-0.39} $ & $   0.31^{+0.07}_{-0.09} $ \\
  3997 &    Im &  73.8 & $  -0.33^{+0.21}_{-0.29} $ & $   0.84^{+0.02}_{-0.02} $ & $   1.09^{+0.22}_{-0.30} $ & $  -1.17^{+0.21}_{-0.29} $ & $  -3.53^{+0.26}_{-0.31} $ & $   0.40^{+0.08}_{-0.10} $ \\
  4036 & SABbc & 122.5 & $   0.39^{+0.13}_{-0.17} $ & $   0.95^{+0.02}_{-0.02} $ & $   1.86^{+0.12}_{-0.13} $ & $  -0.56^{+0.13}_{-0.17} $ & $  -3.32^{+0.14}_{-0.18} $ & $  -0.02^{+0.10}_{-0.10} $ \\
  4107 &    Sc &  88.7 & $   0.27^{+0.13}_{-0.16} $ & $   1.11^{+0.03}_{-0.03} $ & $   1.48^{+0.15}_{-0.16} $ & $  -0.84^{+0.13}_{-0.17} $ & $  -3.41^{+0.15}_{-0.18} $ & $   0.29^{+0.05}_{-0.05} $ \\
  4256 &  SABc & 134.3 & $   0.99^{+0.12}_{-0.16} $ & $   1.39^{+0.03}_{-0.04} $ & $   2.06^{+0.22}_{-0.21} $ & $  -0.40^{+0.12}_{-0.16} $ & $  -3.26^{+0.17}_{-0.19} $ & $  -0.23^{+0.06}_{-0.06} $ \\
  4368 &   Scd & 140.7 & $   0.40^{+0.15}_{-0.20} $ & $   0.92^{+0.01}_{-0.02} $ & $   1.32^{+0.21}_{-0.27} $ & $  -0.52^{+0.15}_{-0.20} $ & $  -3.00^{+0.21}_{-0.22} $ & $   0.36^{+0.08}_{-0.11} $ \\
  4380 &   Scd &  70.5 & $  -0.17^{+0.18}_{-0.24} $ & $   0.88^{+0.02}_{-0.02} $ & $   1.02^{+0.29}_{-0.45} $ & $  -1.05^{+0.18}_{-0.24} $ & $  -3.38^{+0.28}_{-0.29} $ & $   0.42^{+0.05}_{-0.05} $ \\
  4458 &    Sa & 114.3 & $   0.26^{+0.12}_{-0.16} $ & $   0.85^{+0.02}_{-0.02} $ & $   2.15^{+0.11}_{-0.11} $ & $  -0.60^{+0.12}_{-0.16} $ & $  -3.50^{+0.14}_{-0.17} $ & $   0.22^{+0.11}_{-0.11} $ \\
  4555 & SABbc &  97.3 & $   0.13^{+0.14}_{-0.17} $ & $   0.98^{+0.02}_{-0.02} $ & $   1.65^{+0.16}_{-0.19} $ & $  -0.84^{+0.14}_{-0.17} $ & $  -3.49^{+0.17}_{-0.19} $ & $   0.31^{+0.08}_{-0.12} $ \\
  4622 &   Scd &  84.8 & $  -0.29^{+0.18}_{-0.22} $ & $   0.75^{+0.02}_{-0.02} $ & $   0.85^{+0.15}_{-0.18} $ & $  -1.04^{+0.18}_{-0.22} $ & $  -3.29^{+0.20}_{-0.24} $ & $   0.37^{+0.07}_{-0.08} $ \\
  6903 &  SBcd & 157.8 & $  -0.15^{+0.13}_{-0.16} $ & $   0.79^{+0.01}_{-0.01} $ & $   1.78^{+0.21}_{-0.38} $ & $  -0.94^{+0.13}_{-0.16} $ & $  -3.65^{+0.22}_{-0.20} $ & $   0.11^{+0.26}_{-0.17} $ \\
  6918 &  SABb & 140.7 & $   1.04^{+0.12}_{-0.16} $ & $   1.43^{+0.04}_{-0.04} $ & $   1.69^{+0.14}_{-0.12} $ & $  -0.39^{+0.13}_{-0.16} $ & $  -3.07^{+0.14}_{-0.17} $ & $   0.23^{+0.09}_{-0.08} $ \\
  7244 &  SBcd &  95.1 & $  -0.35^{+0.19}_{-0.25} $ & $   0.83^{+0.02}_{-0.02} $ & $   1.40^{+0.18}_{-0.21} $ & $  -1.18^{+0.19}_{-0.25} $ & $  -3.70^{+0.22}_{-0.27} $ & $   0.28^{+0.09}_{-0.14} $ \\
  8196 &    Sb &  92.9 & $   0.78^{+0.12}_{-0.16} $ & $   0.72^{+0.02}_{-0.02} $ & $   2.36^{+0.10}_{-0.10} $ & $   0.06^{+0.12}_{-0.16} $ & $  -2.95^{+0.13}_{-0.17} $ & $  -0.12^{+0.10}_{-0.10} $ \\
  9177 &   Scd &  84.8 & $   0.51^{+0.17}_{-0.21} $ & $   0.77^{+0.02}_{-0.02} $ & $   1.63^{+0.18}_{-0.19} $ & $  -0.26^{+0.17}_{-0.21} $ & $  -2.90^{+0.19}_{-0.23} $ & $   0.23^{+0.12}_{-0.13} $ \\
  9837 &  SABc & 109.2 & $   0.02^{+0.14}_{-0.18} $ & $   1.03^{+0.01}_{-0.02} $ & $   0.40^{+0.50}_{-10.0} $ & $  -1.01^{+0.14}_{-0.18} $ & $  -3.19^{+0.35}_{-10.0} $ & $   0.45^{+0.06}_{-0.06} $ \\
  9965 &    Sc &  73.8 & $   0.39^{+0.13}_{-0.17} $ & $   1.12^{+0.03}_{-0.03} $ & $   1.17^{+0.25}_{-0.38} $ & $  -0.74^{+0.14}_{-0.17} $ & $  -3.14^{+0.22}_{-0.22} $ & $   0.36^{+0.06}_{-0.06} $ \\
 12391 &  SABc & 111.7 & $   0.12^{+0.13}_{-0.17} $ & $   0.98^{+0.02}_{-0.03} $ & $   1.51^{+0.15}_{-0.18} $ & $  -0.86^{+0.13}_{-0.17} $ & $  -3.44^{+0.16}_{-0.18} $ & $   0.24^{+0.06}_{-0.09} $
\enddata
\tablenotetext{a}{Hubble types are taken from Section 5.1 of \citetalias{2010ApJ...716..198B}, which are based on the UGC and RC3 catalogs.}
\tablenotetext{b}{The homogenized RC3 (blue) values, corrected for extinction and inclination, from NED.}
\label{tab:data}
\end{deluxetable*}


In the future, we plan to flux calibrate our H$\alpha$ spectroscopy and
calculate spatially resolved star-formation rates, using our 24$\mu$m data to
correct for dust-enshrouded star formation.  However, in this work, we use the
integrated 21cm continuum luminosity, $L_{21}$, to produce global star-formation
rates, $\sfr$, based on the calibration provided by \citet{2001ApJ...554..803Y}.
\citeauthor{2001ApJ...554..803Y} produced this calibration by matching the local
21cm luminosity density to the local star-formation density.  The simple linear
relationship provided by \citet[][see their equation 13]{2001ApJ...554..803Y} is
sufficient for our study.  However, we note that the $L_{21}$-$\sfr$ calculation
has been improved to account for the systematic underestimate of the
star-formation rate in low luminosity galaxies by \citet{2003ApJ...586..794B}.
The luminosity range of our galaxies is such that the adoption of the
\citet[][see his equation 6]{2003ApJ...586..794B} relation yields a maximum
difference of a factor of two at the low luminosity end and has a negligible
effect on our conclusions.

We calculate $L_{21}$ using the distance, $D$, and $S_{21}$, and then we
calculate $\log(\sfr) = \log(L_{21}) + Z_{\rm SFR}$ using the zero-point,
$Z_{\rm SFR}$ derived by \citet{2001ApJ...554..803Y}.  We assume galaxies follow
the Hubble flow such that the distance is $D = (V_{\rm sys} - V_{\rm pec})/H_0$,
where $V_{\rm pec}$ is the peculiar velocity taken from NED\footnote{
The NASA/IPAC Extragalactic Database, operated by the Jet Propulsion Laboratory,
California Institute of Technology, under contract with the National Aeronautics
and Space Administration; \url{http://nedwww.ipac.caltech.edu/}.
} \citep{2000ApJ...529..786M}.  We calculate $\sfr$ for each sample of $V_{\rm
sys}$ in the dynamical model and use normal distributions for the other
parameters in the calculation ($S_{21}$, $V_{\rm pec}$, $H_0$, and $Z_{\rm
SFR}$) with a known or adopted error.  Finally, similar to
\citetalias{1998ApJ...498..541K}, we calculate an ``effective'' star-formation
rate surface density, $\Ssfre = \sfr/\pi R_{25}^2$, where $R_{25}$ is based on
the diameter at which $\mu_B = 25$ \muu, $D_{25}$, from NED (Table
\ref{tab:data}).

For comparison with $\Ssfre$, we calculate ``effective'' mass surface densities
for each component $j$,
\begin{equation}
\Sigma_{e,j} = \frac{2}{R_{25}^2} \int_0^{R_{25}} R \Sigma_j dR.
\label{eq:avsig}
\end{equation}

We also calculate an effective star-formation efficiency (SFE), $\Ssfre/\Sge$,
and the quantity $\Ssfre\Sge^{-1}\Sse^{-1/2}$.  The quantities used in our
discussion are based on the median (50\% growth) and 68\% confidence interval
(the difference between 16\% and 84\% growth) of the marginalized distributions
for each quantity.  We have inspected the covariance among different parameters
in our posterior distribution; however, these are not discussed further because
they do not influence the conclusions we draw in this paper.  The results of our
analysis are provided in Table \ref{tab:data}.


\begin{figure}
\epsscale{0.9}
%
\plotone{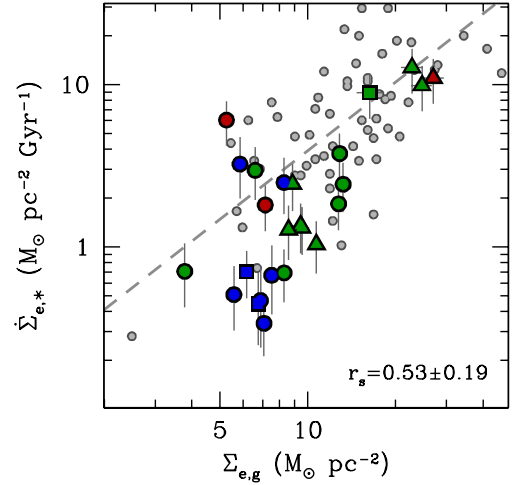}
\caption{
Relation between $\Ssfre$ and $\Sge$ for 25 galaxies from the DiskMass Survey.
Green points are ``bc'' and ``c'' type spirals, with red and blue points being
earlier and later types, respectively.  Circles, triangles, and squares are
unbarred (S), weakly barred (SAB), and barred (SB) galaxies, respectively.  We
provide the rank correlation coefficient, $r_s$, for our sample in the
bottom-right corner.  Gray points are the ``Normal Spirals'' from
\citetalias{1998ApJ...498..541K} and the gray dashed line is the nominal KS law.
Both the data and fit from \citetalias{1998ApJ...498..541K} have been offset to
account for the factor of 1.4 we use to obtain the total gas mass.
}
\label{fig:ks}
\end{figure}


\subsection{Comparison with \citetalias{1998ApJ...498..541K}}
\label{sec:KS}

We compare our sample to the ``Normal Spirals'' presented by
\citetalias{1998ApJ...498..541K} in Figure \ref{fig:ks}, where we have accounted
for the factor of 1.4 difference between our calculation of $\Sge$ and the
hydrogen-only surface densities calculated by
\citetalias{1998ApJ...498..541K}.\footnote{
The value of $\xco$ from \citetalias{1998ApJ...498..541K}, $\xco = 2.8 \times
10^{20}$ cm$^{-2}$ (K \kms)$^{-1}$, is only 4\% different from our own.
}  Our galaxies are roughly consistent with the scatter seen in the
\citetalias{1998ApJ...498..541K} sample.  Additionally, those galaxies with
$\Ssfre < 1 \msol$ pc$^{-2}$ Gyr$^{-1}$ (approximately one-third of our sample)
are consistent with the well-known drop in $\Ssfre$ with respect to the nominal
KS law at low $\Sge$ \citep[e.g.,][Figure 15]{2008AJ....136.2846B}.  Most
important to this comparison, our galaxy sample shows that $\Ssfre$ and $\Sge$
are correlated.

We characterize the correlation between two quantities using the Spearman
rank-order correlation coefficient \citep[$r_s$, see Section 14.6.1 of][]{NR3}.
Exact (anti-)correlation yields $r_s=(-)1$.  By using ranks, $r_s$ is
independent of whether or not one considers the logarithmic or linear
distribution of the data.  Additionally, $r_s$ benefits over the linear (or
Pearson) correlation coefficient because each datum (rank) is drawn from a known
probability distribution, leading to a more straight-forward interpretation of
the significance of the correlation as quantified via the $p$-value \citep{NR3}.
We estimate the error in the correlation coefficient using $10^3$ bootstrap
simulations.  The value of $r_s$ for our $\Ssfre$ and $\Sge$ measurements is
provided in Figure \ref{fig:ks}.  

Although the correlation coefficient measured between $\Sge$ and $\Ssfre$ is
rather significant for our data (the $p$-value rejects the null hypothesis ---
no correlation --- at better than 99\% confidence), the robustness of the
measurement is rather low (with $r_s$ being less than three times its error).
The sample of ``Normal Spirals'' from \citetalias{1998ApJ...498..541K} exhibit a
much stronger correlation, both in terms of significance and robustness, with
$r_s = 0.66\pm0.07$; however, the calculated $r_s$ for the two samples are
consistent within the errors.  If we fit a Schmidt relation to our data, we find
a power-law slope that is within the error of the KS law, but we find a
significantly different normalization.  This can be attributed to the fact that
approximately 80\% of the galaxies in our sample fall below the nominal KS law.
Throughout this paper, we therefore prefer to discuss the correlation between
two quantities, as opposed to fitting regressions.  We work under the assumption
that the correlations we measure should be within the error of those found for
larger samples.  We will fit parameterized forms to our data --- incorporating
independent errors along both axes using {\tt Fitexy} from Section 15.3 from
\citet{NR3} --- when useful for comparing our data with a previous result or
prediction from the literature, but these results are provided primarily for
illustration purposes.


\begin{figure*}
\epsscale{1.15}
%
\plotone{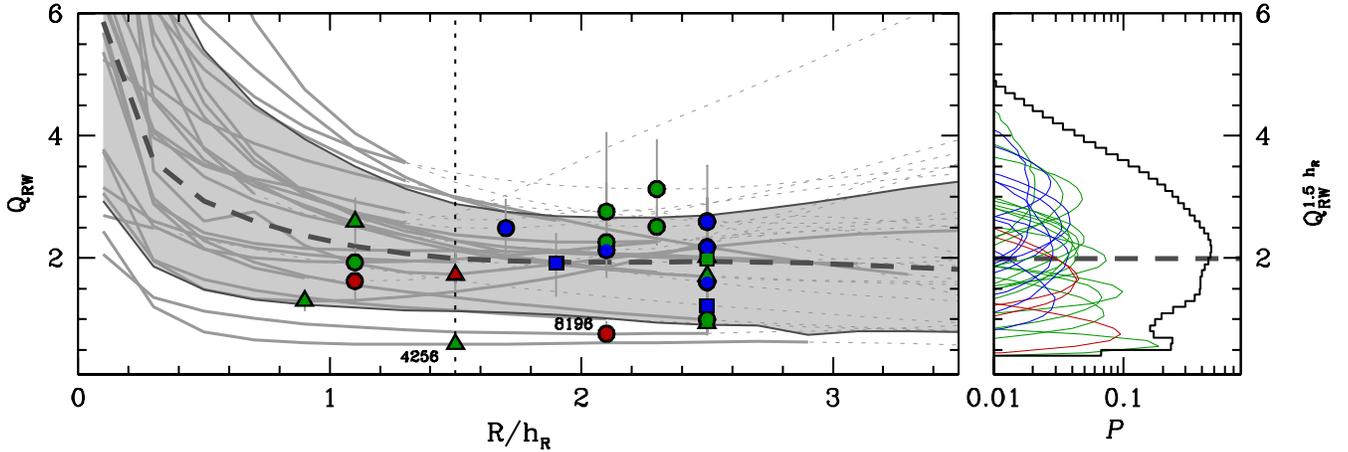}
\caption{
{\it Left} --- The multi-component disk stability parameter, $\qrw$, as a
function of $R/h_R$, determined by the median of the posterior from our
dynamical model of each galaxy.  Regions directly constrained by $\sigma_\ast$
are shown as solid gray lines, whereas regions beyond this (dotted gray lines)
are either constrained by the rotation curves or extrapolations of the model.
The minimum value between $0.1\leq R/h_R\leq2.5$, $\qrwmin$, is marked for each
galaxy; we have labeled the results for UGC 4256 and 8196.  Colors and symbol
types are the same as in Figure \ref{fig:ks}.  The gray dashed line is the
median of the marginalized distributions of $\qrw$ versus $R/h_R$ and the
light-gray region shows the 68\% confidence interval.  The vertical dotted black
line marks $R=1.5h_R$.  {\it Right} --- The probability distribution of $\qrw$
at 1.5 $h_R$ for each galaxy, color coded by Hubble type as in Figure
\ref{fig:ks}.  The black histogram is the probability distribution of $\qrw$
marginalized over all galaxies, with the integral normalized to unity; the gray
dashed line is the median of this distribution.  All other probability
distributions have been normalized relative to their contribution to the total
marginalized probability.
}
\label{fig:combined_q}
\end{figure*}


\subsection{ Stability Results }
\label{sec:result}

We calculate $\qrw$ as a function of radius (in units of the scale length,
$h_R$) for 25 of the 27 galaxies in our sample as shown in Figure
\ref{fig:combined_q}.  The median radial profile of $\qrw$ marginalized over all
galaxies is high near the center (largely driven by the epicyclic frequency) and
asymptotes to a nearly constant value beyond $\sim$1 $h_R$.

Two galaxies --- UGC 4256 and UGC 8196 --- exhibit $\qrw < 1$ over the majority
of their disks, which is difficult to interpret.  Either the dynamical
assumptions made by our modeling approach have led to a value of $\qrw$ that is
systematically in error or the disks of these two galaxies are, in actuality,
unstable according to the criterion derived by \citet{2011MNRAS.416.1191R}.  In
the case of UGC 4256, it might be reasonable to expect the latter because it has
a rather massive molecular component (as measured by its 24$\mu$m surface
brightness) relative to its stellar disk \citepalias{2013A&A...557A.131M}.
Also, UGC 4256 likely suffered a recent interaction and it exhibits a one-armed,
asymmetric morphology.  However, these latter two observations contradict the
assumptions made by our dynamical model, such that we might expect that the low
$\qrw$ values are systematically in error.  In the case of UGC 8196, the galaxy
is bulge-dominated and an outlier in our maximality analysis with an overly
massive baryonic disk; see \citet[hereafter
\citetalias{2011ApJ...739L..47B}]{2011ApJ...739L..47B} and additional discussion
of this galaxy in \citetalias{2013A&A...557A.131M}.  We continue to consider the
results for these two galaxies below, but advise the reader to keep these
caveats in mind.

For the remainder of the paper, we focus on two fiducial measurements of the
stability level.  First, we determine the minimum over the range $0.1\leq
R/h_R\leq2.5$, $\qrwmin$.  For 12 galaxies, the value of $\qrwmin$ is taken at
$2.5h_R$ because $\qrw$ continues to decrease beyond this radius such that the
true minimum of $\qrw$ is not well constrained by the data.  The determined
values of $\qrwmin$ are shown in Figure \ref{fig:combined_q} and provided in
Table \ref{tab:data}.  Second, we calculate $\qrw$ at 1.5 scale lengths,
$\qrw^{1.5h_R}$, which is expected to be an optimal disk value, well away from
any bulge component and often within the radial regime of our stellar velocity
dispersion measurements.  The conclusions we reach in Section \ref{sec:sfrcorr}
are very similar when considering either $\qrwmin$ or $\qrw^{1.5h_R}$; however,
the correlations with $\qrwmin$ are strongest.  Additionally, we are motivated
to use $\qrwmin$ such that we can more directly compare our measurements to the
theoretical results from \citet{2006ApJ...639..879L}.

At $R=1.5h_R$, we find a full range of $0.6 < \qrw^{1.5h_R} < 3.4$ and a
marginalized median of $2.0\pm0.9$.  This value for the disk stability parameter
is close to the values often found in N-body simulations of galaxy disks
\citep[e.g.,][]{2012MNRAS.426.2089R}.  However, we emphasize that this is a {\em
two-component} stability level with corrections for disk thickness, whereas the
majority of N-body simulations consider only the collisionless, single-component
criterion from \citet{1964ApJ...139.1217T}.  The stability parameter in equation
\ref{eq:q} can be corrected such that it is valid for a single-component
collisionless stellar system by calculating $\mathcal{Q}_\ast = \pi Q_\ast/3.36$
\citep{1964ApJ...139.1217T}.  At 1.5$h_R$, we find marginalized median values of
$Q_g = 2.0^{+1.1}_{-0.7}$ and $\mathcal{Q}_\ast = 2.9^{+5.0}_{-1.8}$ with ranges
of $0.5\leq Q_g\leq 4.7$ and $0.5 \leq \mathcal{Q}_\ast \leq 25$.  We also find
that $Q_g < \mathcal{Q}_\ast$ for approximately 65\% of our sample, such that
the calculation of $\qrw^{1.5h_R}$ is often dominated by the contribution of the
gaseous component.

Our marginalized median value of $\langle \qrw^{1.5h_R}\rangle = 2.0\pm0.9$
agrees well with other empirical assessments.  In particular, \citet[][see their
Figure 5]{2013MNRAS.433.1389R} studied the stability parameter in the galaxy
sample presented by \citet{2008AJ....136.2782L} and found a median value of
$\qrw \sim 2$ over most of the optical radius.  However, they also concluded
that it is the stellar component that typically dominates the calculation of the
two-component $\qrw$, whereas we find that it is the gas component that most
often dominates.  It is more likely that this difference is due to our different
analysis methods rather than an intrinsic difference in the galaxy samples.

In terms of Hubble type, our galaxies range from Sa to Im (Table \ref{tab:data})
and the \citet{2008AJ....136.2782L} sample has a comparable range from Sab to Im
type.  However, our galaxies are more strongly concentrated toward Sc and Scd
types, whereas the \citet{2008AJ....136.2782L} is evenly distributed between Sb
and Sd with a peak at Im types.

With respect to the stability level of the stellar component,
\citet{2013MNRAS.433.1389R} adopted the stellar mass surface densities and
velocity dispersions calculated by \citet{2008AJ....136.2782L}.
\citeauthor{2008AJ....136.2782L} assumed a $K$-band mass-to-light ratio of
$\Upsilon_K = 0.50\ \msol/\lksol$, an oblateness relation of $h_R/h_z = 7.3$,
and an isothermal disk ($k=2.0$) to calculate $\sigma_R$ using the adopted
stellar surface density and $\alpha = 0.6$.  The most significant difference
with respect to our approach is that, as shown in Section 3.4 of
\citetalias{2013A&A...557A.131M}, our {\em dynamical} calculations of the
surface mass density yield a mean value of $\langle\Upsilon_K\rangle = 0.31\
\msol/\lksol$.  In total, we expect the approach of \citet{2013MNRAS.433.1389R}
leads to values of $\mathcal{Q}_\ast$ that are, on average, a factor of 0.8
times our own.

A more significant difference is in the calculation of $Q_g$.  Our analysis
assumes $\Sigma_g = 1.4\Sigma_H$, $\sigma_{cg} = \sigma_{ig}/2$ (leading to a
mean value of 8.3 \kms), and $\xco = 2.7 \times 10^{20}$ cm$^{-2}$ (K
\kms)$^{-1}$; however, \citet{2008AJ....136.2782L} adopt $\Sigma_g = \Sigma_H$,
$\sigma_{cg} = 11$ \kms, and $\xco = 2.0 \times 10^{20}$ cm$^{-2}$ (K
\kms)$^{-1}$.  The differences between our calculation and that from
\citet{2013MNRAS.433.1389R} are at their extrema when one assumes the gas is
either fully atomic or fully molecular, such that their calculations of $Q_g$
should be, on average, factors of 1.9--2.5 times our own.  This is likely why we
find more galaxies with $Q_g<\mathcal{Q}_\ast$ than in the two-component
analysis from \citet[][see also
\citealt{2011MNRAS.416.1191R}]{2013MNRAS.433.1389R}.

We are confident in our calculations of $\mathcal{Q}_\ast$ due to our reliance
on stellar kinematic data, as opposed to an inferred $\Upsilon_K$; yet our
calculations of $Q_g$ \citep[as well as those produced
by][]{2013MNRAS.433.1389R} do, unfortunately, depend on assumed factors.  All of
our assumptions are justified; however, our analysis would benefit from more
direct constraints on the gas-phase metallicity, $\sigma_{cg}$, and $\xco$ for
each galaxy.


\begin{figure}
\epsscale{1.1}
%
\plotone{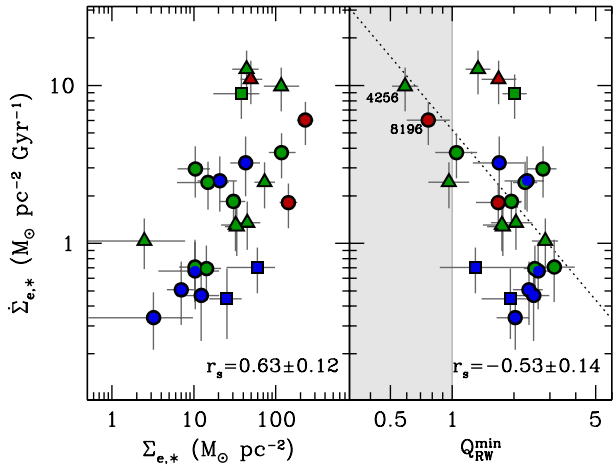}
\caption{
The effective star-formation-rate surface density, $\Ssfre$, as a function of
effective stellar mass surface density, $\Sse$, and the minimum disk stability
parameter, $\qrwmin$.  Colors and symbol types are the same as in Figure
\ref{fig:ks}.  The light-gray region denotes nominal disk instability.  We
provide $r_s$ in the bottom-right corner of each panel.  The two galaxies with
$\qrw<1$ over the majority of their disks (see Figure \ref{fig:combined_q}) are
labeled in the right panel.  The dotted line is the best fitting power-law
relationship with a fixed power-law slope of -1.54 \citep{2006ApJ...639..879L}.
Our data appear to show a much steeper relationship (with a power-law slope of
approximately -3).
}
\label{fig:sigsfr}
\end{figure}


\section{ (Anti-)Correlations with Star-Formation Rate }
\label{sec:sfrcorr}

In this section, we explore correlations between $\Ssfre$, $\Sse$, and
$\qrwmin$.  We are the first to explore these correlations using measurements of
stellar mass surface density and disk stability levels that are directly
constrained by stellar kinematics.

\begin{figure}
\epsscale{1.1}
%
\plotone{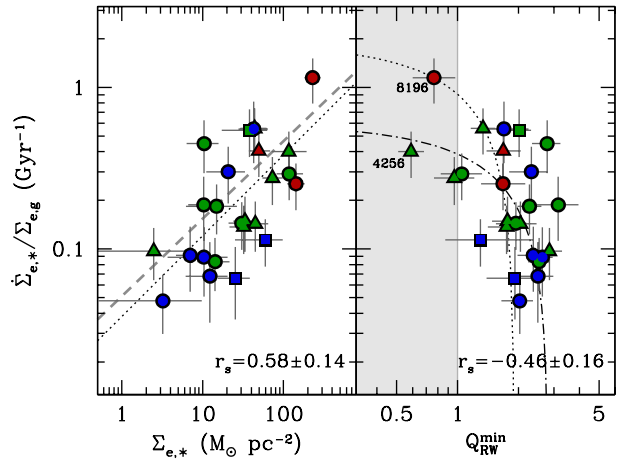}
\caption{
The star-formation efficiency (SFE), $\Ssfre/\Sge$, as a function of effective
stellar mass surface density, $\Sse$, and minimum disk stability parameter,
$\qrwmin$.  Colors, symbol types, and the light-gray region are the same as in
Figure \ref{fig:sigsfr}.  In the left panel, the dotted line is the best-fitting
power law with a power-law slope of 0.5, showing consistency with the
expectation of \citetalias{2010ApJ...721..975O}; the gray dashed line is the
relation from \citet{2011ApJ...733...87S}.  In the right panel, the dotted line
shows the best-fitting {\em linear} relationship with a slope of -1.0
\citep{2006ApJ...639..879L}, whereas the dot-dashed line shows the best fit when
both the slope and intercept are free parameters.  Both panels exhibit a
correlation; however, the correlation between $\qrwmin$ and the SFE is rather
weak.
}
\label{fig:sigsfg}
\end{figure}

Figure \ref{fig:sigsfr} demonstrates that our data show significant and robust
correlations of $\Sse$ and $\qrwmin$ with $\Ssfre$; however, both panels in
Figure \ref{fig:sigsfr} exhibit large scatter, with a range of up to 1.5 dex
seen in $\Ssfre$ at fixed $\Sse$ or fixed $\qrwmin$.  By comparison, this range
is reduced to approximately 1 dex when considering $\Ssfre/\Sge$ (the SFE, see
Figure \ref{fig:sigsfg}) instead.  Nonetheless, the correlation between $\Sse$
and $\Ssfre$ shown in Figure \ref{fig:sigsfr} is the strongest and most robust
of those presented in this paper.  We find $\Ssfre$ is anti-correlated with
$\qrwmin$ at better than 99\% confidence considering both the null hypothesis
and the error in $r_s$; however, neither gas-only nor star-only stability
calculations exhibit such a correlation.  Thus, we find that in relating disk
stability levels to star formation, it is important to incorporate both
components, gas and stars, in the stability assessment.

Based on GADGET N-body simulations of gaseous disk galaxies (using smoothed
particle hydrodynamics), \citet{2006ApJ...639..879L} find $\Ssfr \propto
[Q_{sg,{\rm min}}(\tau_{\rm SF})]^{-1.54}$.  In their analysis, they adopt the
stability parameter derived by \citet{2001MNRAS.323..445R}, using the wavenumber
of the perturbation that yields the minimum two-component (gas$+$stars) disk
stability level, and they refer to this as $Q_{sg}$.
\citet{2013MNRAS.433.1389R} have shown that $\qrw$ is a good approximation to
this usage of the \citet{2001MNRAS.323..445R} formulation, but without the need
to determine the minimizing wavenumber.  The value $Q_{sg,{\rm min}}(\tau_{\rm
SF})$ provided by \citet{2006ApJ...639..879L} is the minimum value of $Q_{sg}$
over all radii at one $e$-folding time of the star-formation rate, $\tau_{\rm
SF}$.  Under the expectation that our galaxies are all quiescently star-forming
and that the evolution of $Q_{sg}$ is moderate at $t>\tau_{\rm SF}$, the above
proportionality from \citet{2006ApJ...639..879L} should be reflected in Figure
\ref{fig:sigsfr}.  A caveat to this comparison is that the galaxy simulations in
\citet{2006ApJ...639..879L} were of largely unstable disks with $Q_{sg,{\rm
min}}(\tau_{\rm SF}) \lesssim 1$.

The dotted line in Figure \ref{fig:sigsfr} is a fit to our data done by fixing
the slope to that expected by \citet{2006ApJ...639..879L}.  Our data appear to
follow a much steeper relation, with a power-law slope that is approximately -3.
However, we note that the correlation seen between $\qrwmin$ and $\Ssfre$
depends rather strongly on the two galaxies with $\qrw < 1$ over most of the
disk.



\begin{figure}
\epsscale{1.1}
%
\plotone{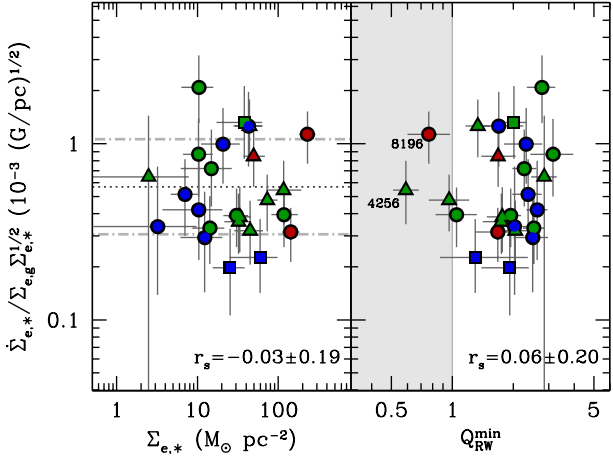}
\caption{
The expected constant of proportionality for $\Ssfre/\Sge \propto \Sse^{1/2}$ as
a function of effective stellar mass surface density, $\Sse$, and minimum disk
stability parameter, $\qrwmin$.  Colors, symbol types, and the light-gray region
are the same as in Figure \ref{fig:sigsfr}.  The dotted line and the gray
dot-dashed lines in the left panels gives the error-weighted geometric mean and
standard deviation, respectively:
$\langle\log(\Ssfre\Sge^{-1}\Sse^{-1/2})\rangle = -3.25\pm0.27$ in units of
$(G/{\rm pc})^{1/2}$.
}
\label{fig:sigsfgs}
\end{figure}


\citet{2006ApJ...639..879L} also predict a linear relationship between
$\Ssfre/\Sge$ and $\qrwmin$.  \citet{2008AJ....136.2782L} found no evidence for
such a correlation using spatially resolved observations.  However, Figure
\ref{fig:sigsfg} shows that our data exhibit a weak correlation between
$\Ssfre/\Sge$ and $\qrwmin$, which is again highly dependent on the two galaxies
with $\qrwmin < 1$.  Figure \ref{fig:sigsfg} shows the best-fitting {\em linear}
relationship with a slope of -1.0 \citep[dotted line; as expected
by][]{2006ApJ...639..879L} and the result after fitting both the slope and
intercept (we find a best-fitting slope of -0.2; dot-dashed line).  Thus, our
data are at least suggestive of the relation expected by
\citet{2006ApJ...639..879L}, albeit qualitatively.

Figure \ref{fig:sigsfg} also shows the correlation of $\Ssfre/\Sge$ with $\Sse$.
Our data are consistent with the empirical result from
\citet{2011ApJ...733...87S}, and with a proportionality derived by
\citetalias{2010ApJ...721..975O} for the outer parts of stellar-dominated,
constant-scale-height disks, $\Ssfre/\Sge \propto \Sse^{1/2}$.  The dotted line
in the left panels of Figures \ref{fig:sigsfg} and \ref{fig:sigsfgs} show the
result of fitting this proportionality to our data.  Given the order of
magnitude range in $\Ssfre\Sge^{-1}\Sse^{-1/2}$, we might expect other
correlations to exist.  However, we find no residual correlations of this
quantity with respect to, e.g., $h_z$, $v_c$, $\kappa$, $\sigma_R$,
$\sigma_{cg}$, $\alpha$, $\Sge$, $\Sse$, $\Sge/\Sse$, or $\qrwmin$ (Figure
\ref{fig:sigsfgs}): the significance of all correlations are low with
$|r_s|<0.2$ and the measurements of $r_s$ are all consistent with no correlation
according to their error.  In units of $(G/{\rm pc})^{1/2}$, we find an
error-weighted geometric mean of $\langle\log(\Ssfre\Sge^{-1}\Sse^{-1/2})\rangle
= -3.25\pm0.27$.\footnote{
$G = 4.30\times10^{-3}$ (km/s)$^2$ pc $\msol^{-1}$
} This result is further discussed in Section \ref{sec:conclusions}.

\section { An Expectation from Scaling Relations }
\label{sec:scaling}

Instead of exploring a theoretical understanding of the correlation between
$\Ssfre$ and $\qrw$ shown in Section \ref{sec:sfrcorr}, we ask two questions.
Given other (selected) ensemble properties of late-type galaxies, should we {\em
expect} an anti-correlation between $\Ssfre$ and $\qrw$?  If so, does that
expectation match our direct measurements?  We address these questions by
building a closed system of empirical scaling relations (i.e., the number of
equations matches the number of unknowns) from which we can compute $\qrw$ for a
given $\Ssfr$ in an idealized galaxy.

\subsection{ Calculation Details }
\label{sec:calcpred}

Equations \ref{eq:q}, \ref{eq:efreq}, and \ref{eq:thickq}--\ref{eq:qweight}
require $v_c$, $\sigma_{cg}$, $\sigma_R$, $\Sigma_g$, and $\Sigma_\ast$ to
produce $\qrw$.  For the calculation, we ignore the effects of any bulge
component and assume a constant $K$-band mass-to-light ratio, $\Upsilon_K$
\citepalias[cf.][]{2013A&A...557A.131M}.  We assume $\sigma_{cg}$ is constant
and we set it directly.  The remaining four quantities are determined by setting
$\sfr$, the disk central surface brightness ($\mu_{0,K}$), $h_R$, $R_{25}$,
$\Upsilon_K$, and $\alpha$ and combining these quantities with a set of
empirical scaling relations as described below.  We emphasize that we are not
suggesting that our seven input parameters are the fundamentally relevant
physical quantities of a galaxy (in the same sense as mass, age, and chemical
composition are relevant to a star); these are simply the quantities that we
need to calculate $\qrw$.

The input parameters set $\Sigma_\ast(R) = \Sigma_{\ast,0}\exp(-R/h_R)$
directly, where $\Sigma_{\ast,0} = \Upsilon_K I_{0,K}$, $\log I_{0,K} =
-0.4(\mu_{0,K} - M_{\odot,K} - 21.57)$, and $M_{\odot,K} = 3.3$ is the absolute
$K$-band magnitude of the sun \citepalias{2011ApJ...742...18W}; $\Upsilon_K$ is
in solar units ($\msol/\lksol$) and $I_{0,K}$ is in units of
$\lksol~\!$pc$^{-2}$.  We calculate $\sigma_R(R) = \sigma_z(R)/\alpha$ using the
input value of $\alpha$ and
\begin{equation}
\sigma_z = (\pi\ k\ G\ h_z\ \Sigma_{\rm dyn})^{1/2}, 
\label{eq:sigz}
\end{equation}
where $\Sigma_{\rm dyn} = \Sigma_\ast + \Sigma_g$ and $k=1.5$ for an exponential
mass distribution \citep{1988A&A...192..117V}; we determine $h_z$ using its
scaling with $h_R$ from \citetalias{2010ApJ...716..234B}.

We parameterize the circular-speed curve by a hyperbolic tangent function that
has two parameters:  the asymptotic rotation speed ($V_{\rm rot}$) and the
radius ($h_{\rm rot}$) at which $v_c = \tanh(1.0)V_{\rm rot} = 0.76 V_{\rm
rot}$.  We have chosen this form so that we can set its parameters based on two
scaling relations.  First, \citet[][Figure 17]{2013ApJ...768...41A} find
\begin{equation}
\log h_{\rm rot} = \log h_R - 0.714 \log \left[\frac{V_{\rm
rot}}{\mbox{\kms}}\right] + 1.47,
\label{eq:hrot}
\end{equation}
which is in good agreement with the scaling relation found by
\citet{2010A&A...519A..47A} but adds a secondary dependence on the rotation
velocity.  Second, we use the relation between disk maximality
($\mathcal{F}^{2.2}_{\rm bary} = V_{\rm bary}/V_c$ at $R=2.2h_R$) and
$\mu_{0,K}$ from \citetalias{2013A&A...557A.131M}: $\mathcal{F}^{2.2}_{\rm bary}
= 0.56 - 0.08 (\mu_{0,K}-18)$.  We calculate the baryonic disk rotation speed,
\begin{equation}
V_{\rm bary} = 0.88 \left( 1 - 0.28\frac{h_R}{h_z}\right) (\pi G \Sigma_{{\rm
dyn},0} h_R)^2,
\label{eq:vbary}
\end{equation}
following \citetalias{2011ApJ...739L..47B}.  Equation \ref{eq:vbary} assumes
that $\Sigma_{\rm dyn}$ follows a pure exponential with a scale length equal to
that of the stars.  However, this is only explicitly true in the limit where
there is no gas disk.  For our calculations, we adopt the rough approximation
\begin{equation}
\Sigma_{{\rm dyn},0} = \Sigma_{g,R}\exp(R/h_R) + \Sigma_{\ast,0},
\end{equation}
where $\Sigma_{g,R}$ is the gas mass surface density at radius $R$ (see below).
We numerically solve for $V_{\rm rot}$ and $h_{\rm rot}$ given the input
$\mu_{0,K}$, $\Upsilon_K$, and $h_R$, and the calculated $\Sigma_{{\rm dyn},0}$.

We use two approaches to define the functional form of $\Sigma_g(R)$, which is
then normalized by the star-formation law (see below).  First,
\citet{2012ApJ...756..183B} found the total hydrogen mass is well described by a
single exponential with an $e$-folding length of $0.61 R_{25}$.  We term this
the ``BB'' approach.  Second, we separate the hydrogen mass into its molecular
and atomic components according to the scaling relation found by
\citet{2011MNRAS.415...32S}:
\begin{equation}
\log \left[\frac{\mhi}{\msol}\right] = 1.01 \left(\log
\left[\frac{\mht}{\msol}\right] + 0.42 -
\log\left[\frac{\xco}{2.0}\right]\right),
\label{eq:h1toh2}
\end{equation}
where $\mhi+\mht$ is the total hydrogen mass and we have generalized the
relation for any (constant) $\xco$ in units of $10^{20}$ cm$^{-2}$ (K
\kms)$^{-1}$; \citetalias{2013A&A...557A.131M} shows equation \ref{eq:h1toh2} is
fully consistent with our galaxy sample.  We assume $\sdmh$ follows an
exponential with the same scale length as the stars \citep{2001ApJ...561..218R},
and $\sdhi$ follows a Gaussian in radius with a center and dispersion of,
respectively, $0.39 \rhi$ and $0.35 \rhi$ \citep{TPKMPhD}.  We calculate $\rhi$
--- the radius at which $\sdhi\sim 1$ \sdu --- using the relation derived by
\citet{2001A&A...370..765V}:
\begin{equation}
\log \left[\frac{\rhi}{\rm kpc}\right] = (\log \left[\frac{\mhi}{\msol}\right] -
7.26) / 1.86.
\label{eq:rh1}
\end{equation}
Finally, we use Brent's minimization method \citep{NR3} to solve the set of
non-linear equations that yield the defining parameters for $\sdhi(R)$ and
$\sdmh(R)$.  We term this the ``MA'' approach.  In both approaches, we set
$\Sigma_g = 1.4 \Sigma_{\rm H} = 1.4(\sdhi+\sdmh)$.

Finally, we use two star-formation laws to determine the normalization of the
hydrogen mass surface density profile.  First, we adopt the KS law from
\citetalias{1998ApJ...498..541K}:
\begin{equation}
\log \left[\frac{\Sigma_{\rm e,H}}{\msol {\rm pc}^{-2}}\right] =  0.71\ \log
\left[\frac{\Ssfre}{\msol {\rm pc}^{-2} {\rm Gyr}^{-1}}\right] + 0.47,
\label{eq:kslaw}
\end{equation}
where $\Sigma_{\rm e,H}$ is the effective hydrogen mass surface density within
$R_{25}$ and we calculate $\Ssfre = \sfr/\pi R_{25}^2$, as done for our data.
Second, we assume
\begin{eqnarray}
\log \left[\frac{\Sigma_{\rm e,H}}{\msol {\rm pc}^{-2}}\right] & = & \log
\left[\frac{\Ssfre}{\msol {\rm pc}^{-2} {\rm Gyr}^{-1}}\right] + 1.15 \nonumber
\\ & & - 0.48 \log \left[\frac{\Sse}{\msol {\rm pc}^{-2}}\right]
\label{eq:eslaw}
\end{eqnarray}
from \citet[][see their Equation 6]{2011ApJ...733...87S}, where we have adjusted
for their factor of 1.36 used to obtain the total gas mass surface density from
the hydrogen mass surface density.  We follow their nomenclature by referring to
this as the ``extended Schmidt'' (ES) law.  The combinatorics of the two
star-formation laws and the two approaches used to distribute the hydrogen mass
lead to four methods for calculating $\qrw$.


\begin{deluxetable}{ r r r r r }
\tabletypesize{\small}
\tablewidth{0pt}
\tablecaption{ $\delta Q$ Quartiles }
\tablehead{ & \multicolumn{4}{c}{Approach} \\ \cline{2-5} \colhead{Growth} & \colhead{KS:MA} & \colhead{KS:BB} & \colhead{ES:MA} & \colhead{ES:BB} }
\startdata
  0.25 &  0.12 & 0.12 & 0.18 & 0.14 \\
  0.50 &  0.21 & 0.24 & 0.23 & 0.23 \\
  0.75 &  0.44 & 0.38 & 0.45 & 0.47 \\
  1.00 &  0.72 & 0.55 & 1.55 & 0.70
\enddata
\label{tab:quart}
\end{deluxetable}



\begin{figure}
\epsscale{1.15}
%
\plotone{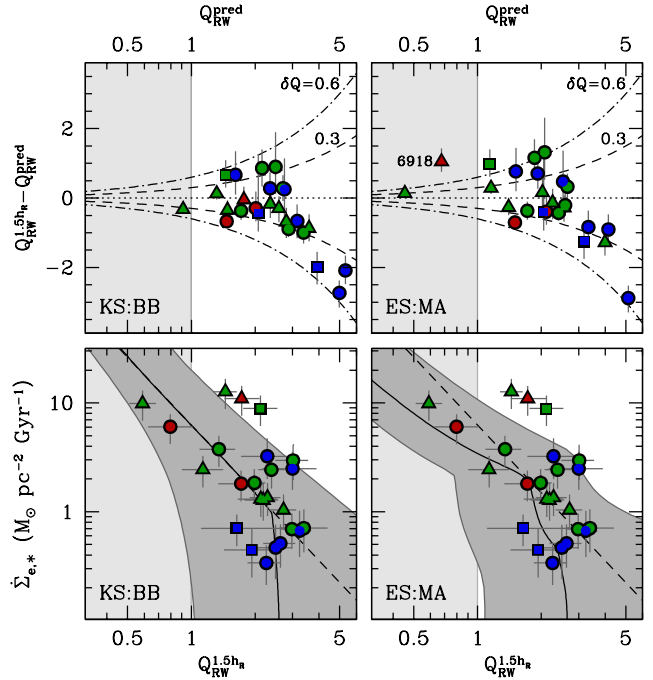}
\caption{
Comparison of predicted $\qrw$ with the measured values at $1.5h_R$; panels to
the left and right use the KS:BB and ES:MA approaches, respectively.  Colors,
symbol types, and the light-gray region are the same as in Figure
\ref{fig:sigsfr}.  The top panels show the difference between the measured
$\qrw^{1.5h_R}$ and the prediction.  The dashed and dot-dashed lines show a
ratio of the two axes equal to $\pm0.3$ and $\pm0.6$, respectively.  The bottom
panels show the measurements of $\Ssfre$ and $\qrw^{1.5h_R}$.  The solid black
line in each panel predicts the correlation between these two quantities
assuming the mean properties of our galaxy sample; the black dashed line, shown
in both panels, is a power-law fit to the KS:BB result between $0.5<\qrw^{\rm
pred}<2.0$.  The dark-gray region is the envelope encompassing all correlations
generated when using the values of $\Upsilon_K$, $\mu_{0,K}$, $h_R$, $R_{25}$,
$\alpha$, and $\sigma_{cg}$ specific to each galaxy.
}
\label{fig:scaling}
\end{figure}


\subsection{ Results }
\label{sec:scaleres}

We first compare our measurements of $\qrw$ from the modeling of our data to the
predictions based on the calculations described above.  Any difference can be
directly attributed to the systematic errors in our simplifying assumptions;
e.g., the assumption for the detailed form of the surface density profiles.  To
focus the following discussion, we compare the data, $\qrw^{1.5h_R}$, and the
prediction, $\qrw^{\rm pred}$, at a single radius, 1.5$h_R$, and define the
quantity $\delta Q=\left|\qrw^{1.5h_R}/\qrw^{\rm pred}-1\right|$.  Our
comparison of the data with the model at 1.5$h_R$ is unimportant to our
conclusions because both the model and the data (Figure \ref{fig:combined_q})
exhibit a slow change in $\qrw$ beyond this radius.

We provide quartiles of $\delta Q$ in Table \ref{tab:quart}, the left-most
column gives the growth of $\delta Q$ and the remaining columns give its value
for the four different approaches to the calculation.  The predicted disk
stability parameters are within 25\% of the measured values for half of our
galaxy sample, regardless of the approach used.  For the entire sample however,
Table \ref{tab:quart} shows the KS:BB and ES:MA methods are, respectively, the
best and worst approaches for predicting $\qrw$.  Interestingly, KS:BB is the
simplest approach and ES:MA is the most complex.  We continue by only comparing
these two approaches and discussing their predictions.

The top two panels of Figure \ref{fig:scaling} show the individual results in
the comparison of $\qrw^{1.5h_R}$ and $\qrw^{\rm pred}$.  Lines are provided
that denote $\delta Q = 0.0$, 0.3, and 0.6.  Thus, we show that the maximum
value for $\delta Q$ when using the ES:MA method is largely due to the outlying
prediction for UGC 6918 (labeled in the top-right panel of Figure
\ref{fig:scaling}).  We also find three late-type spirals that have a rather
large $\qrw^{\rm pred}$ for the KS:BB method; this is likely due to their
significantly larger value of $\Sge$ than that predicted by the KS law (see
Figure \ref{fig:ks}).  Despite these caveats, our prediction of $\qrw^{1.5h_R}$
is reasonable in either approach with $\delta Q\approx0.35$ at 68\% growth.

The bottom two panels of Figure \ref{fig:scaling} compare the measured and
predicted trend of $\qrw$ with $\Ssfre$.  Our combination of empirical scaling
relations {\em predict} an anti-correlation between $\Ssfre$ and $\qrw$:
Adopting the (unweighted) average properties of our sample --- $\sigma_{cg}=8.3$
\kms, $\mu_{0,K}=17.8$ \muu, $h_R=4.3$ kpc, $R_{25}=18.0$ kpc, $\Upsilon_K=0.36
\msol/\lksol$, and $\alpha=0.57$ --- we vary $\sfr$ to produce the black line in
each of the bottom panels of Figure \ref{fig:scaling}.  However, the behavior of
the two methods is different.

Both methods exhibit an inflection in the trend at $\qrw^{\rm pred}> 2$, which
occurs because the calculation of $\qrw$ transitions to the regime where
$\mathcal{T}_\ast \qstar < \mathcal{T}_g \qgas$ at higher $\qrw^{1.5h_R}$ (see
Equation \ref{eq:qrw}).  For $\qrw^{\rm pred}< 2$, the KS:BB approach yields a
well-behaved relationship between $\Ssfre$ and $\qrw^{\rm pred}$ due to its
simple description of the gas distribution:  a power-law slope of -2.07 is an
excellent fit to the KS:BB results between $0.5<\qrw^{\rm pred}<2.0$ (as shown
by the dashed line).

The behavior of the ES:MA method is more complex, exhibiting a second inflection
of the trend at $\qrw^{\rm pred}<2$.  This inflection occurs because, as
$\Ssfre$ increases, the dependence of the \hone\ distribution on its total mass
results in a transition from an \hone-dominated $\Sigma_g$ to an \molh-dominated
$\Sigma_g$ at $1.5h_R$.  The dependence of the \hone\ distribution on its total
mass in the MA approach significantly contributes to its larger $\delta Q$.  The
MA approach suggests two-thirds of the galaxies have $\rhi/R_{25} < 1$, whereas
our direct measurements show $\rhi/R_{25} > 1$ for {\em all} galaxies
\citep{TPKMPhD}.  This failing is due to the low $\Sge$ predicted by the ES and
KS laws at low $\Ssfre$, not due to our adoption of equation \ref{eq:rh1}.  The
$\mhi$-$\rhi$ relation is very tight such that, when considered as an isolated
sample, our galaxies exhibit a very similar scaling relation \citep{TPKMPhD}.

Instead of adopting the mean properties of our sample, we have also calculated
the trend of $\qrw^{\rm pred}$ with $\Ssfre$ when adopting the specific
properties of each galaxy.  We represent the range in the resulting trends using
the dark-gray envelope about the black line in the lower two panels of Figure
\ref{fig:scaling}.  Galaxies should populate this region in so far as our galaxy
sample is representative of the range of $\Upsilon_K$, $\mu_{0,K}$, $h_R$,
$R_{25}$, $\alpha$, and $\sigma_{cg}$ for late-type galaxies
\citepalias{2010ApJ...716..198B} and considering the systematic errors in
$\qrw^{\rm pred}$.  Therefore, our results show that one should {\em expect} a
correlation between $\qrw$ and $\Ssfre$ in a galaxy population, particularly at
$\qrw^{1.5h_R} < 2$.

\section{ Summary \& Discussion }
\label{sec:conclusions}

\subsection{ Summary of Analyses and Empirical Findings }

In this paper, we have used data from the DiskMass Survey, briefly described in
Section \ref{sec:data}, to study the relationship between dynamical properties
of galaxy disks and their star-formation activity.  Unlike other local-universe
surveys, the DiskMass Survey has directly measured the stellar kinematics in a
sample of galaxy disks, which are critical to dynamical calculations of both
stellar surface mass density and disk stability level.  Our calculations are
based on a generative --- fully probabilistic --- model of the relevant gaseous
and stellar observations, adopting a simple analytical model for the disk
dynamics.  We introduce our probabilistic modeling approach in Section
\ref{sec:genmodel}, we briefly outline the dynamical model in Appendix
\ref{app:dynmodel}, and we discuss our usage of a sophisticated MCMC algorithm
to sample the probabilistic model in Appendix \ref{app:converge}.  A more
complete discussion of our modeling approach will be the subject of a
forthcoming paper.

We calculate the star-formation rate, $\sfr$, for each of our galaxies using
measurements of the 21cm radio continuum \citep[NVSS;][]{TPKMPhD} and the
calibrations from \citet{2001ApJ...554..803Y} (see Section \ref{sec:qeqns}).
These calculations are performed by including appropriate probability
distributions for the relevant quantities in our probabilistic model (equation
\ref{eq:pfull}).  We calculate an effective star-formation-rate surface density,
$\Sigma_{\rm SFR}\equiv\Ssfre$, where the surface area of the disk is determined
by $R_{25}$ measurements from NED (Table \ref{tab:data}).  Combined with
effective gas mass surface densities from our probabilistic model (equation
\ref{eq:avsig}), we compare our galaxy sample to the ``Normal Spirals'' from
\citetalias{1998ApJ...498..541K} in Section \ref{sec:KS} (see Figure
\ref{fig:ks}).  We find that the galaxies in our sample have low star-formation
rates relative to the KS law, but are fully consistent with the low
star-formation end of the data presented by \citet{2008AJ....136.2846B}.

We calculate the disk stability parameter derived by
\citet{2011MNRAS.416.1191R}, $\qrw$, which incorporates both the gaseous and
stellar components (see Section \ref{sec:qeqns}) and includes corrections for
the disk thickness.  In Section \ref{sec:result}, we find a stability parameter
of $\qrw=2.0\pm0.9$ at 1.5 $h_R$ marginalized over all galaxies in our sample.
This result is comparable to other empirical assessments from the literature.
In particular \citet{2013MNRAS.433.1389R} similarly find $\qrw\sim2$ for their
sample; however, they also find that the stellar component most often dominates
the disk stability level.  This is contrary to our results, which show that the
gas-only stability parameter is lower than the star-only stability parameter
($Q_g < \mathcal{Q}_\ast$) for 65\% of our sample.  These different findings are
more likely because of differences in the data used and the detailed analysis
methods, as described in Section \ref{sec:result}, not because of an intrinsic
difference in our galaxy samples.

A stability parameter of $\qrw\sim2$ is also comparable to expectations from
N-body simulations; however, it should be noted that these theoretical studies
typically calculate the nominal (infinitely thin, single-component,
collisionless) \citet{1964ApJ...139.1217T} criterion ($\mathcal{Q}_\ast$), as
opposed to our calculations for a multi-component, non-zero thickness disk.  An
important avenue for numerical simulations of galaxy disks is to study the
stability levels of disks that include realistic gas components.

The primary goal of this paper has been to explore any dependence of the
star-formation rate on two dynamical properties of our sample, $\Sse$ and $\qrw$
(see Section \ref{sec:sfrcorr}).  Our primary findings are:
\begin{itemize}
\item  There is a clear correlation between $\Sse$ and $\Ssfre$ and a clear
anti-correlation of $\qrwmin$ with $\Ssfre$ (Figure \ref{fig:sigsfr}).
\item The anti-correlation between $\qrwmin$ and $\Ssfre$ is expected given a
theoretical study by \citet{2006ApJ...639..879L}.  However, our galaxies exhibit
significantly higher stability levels than in their simulations and our data
show a steeper power-law slope in the relation.
\item We find that the star-formation efficiency (SFE; $\Ssfre/\Sge$) is
correlated with $\Sse$ (Figure \ref{fig:sigsfg}), which is expected both
observationally \citep{2011ApJ...733...87S} and theoretically
\citepalias{2010ApJ...721..975O}.  In detail, our data are consistent with the
proportionality $\Ssfre/\Sge \propto \Sse^{1/2}$, which is a limiting behavior
of the theory derived by \citetalias{2010ApJ...721..975O}; see also
\citet{2011ApJ...743...25K, 2013ApJ...776....1K}.
\item If star formation in our galaxy sample is not strongly affected by other
physical properties, the quantity $\Ssfre \Sge^{-1} \Sse^{-1/2}$ should be
roughly constant.  Indeed, we find that this quantity is effectively
uncorrelated with the large number of physical quantities we have calculated
(listed in Section \ref{sec:sfrcorr}; see the examples of $\Sse$ and $\qrwmin$
in Figure \ref{fig:sigsfgs}).  However, the scatter in the data is large.  We
find an error-weighted geometric mean of $\langle\log(\Ssfre
\Sge^{-1}\Sse^{-1/2})\rangle = -3.25\pm0.27$ in units of $(G/{\rm pc})^{1/2}$
and a range of $-3.7\leq\log(\Ssfre \Sge^{-1}\Sse^{-1/2})\leq-2.7$.
\item Although the correlation is weak, we also find an indication that the SFE
is anti-correlated with $\qrwmin$ (Figure \ref{fig:sigsfg}).  This contradicts
previous observational studies \citep{2008AJ....136.2782L}, but is roughly
consistent with the expectation provided by \citet{2006ApJ...639..879L}.
\end{itemize}

The anti-correlation between star-formation rate and disk stability parameter
from Figure \ref{fig:sigsfr} is {\em not} seen if we instead consider the
gas-only ($Q_g$) or star-only ($\mathcal{Q}_\ast$) stability parameters.  In
terms of the effect on star formation, this result is reasonable in that neither
$Q_g$ nor $\mathcal{Q}_\ast$ includes the gravitational effects of the other
component.  Thus, our results show the importance of considering {\em both}
components in assessing the effect of the disk stability level on star
formation.

In Section \ref{sec:scaling}, we quantitatively predict $\qrw$ at 1.5$h_R$ for
the galaxies in our sample based on a closed system of scaling relations and the
following global quantities for our galaxies: $\sfr$, $\Upsilon_K$, $\mu_{0,K}$,
$h_R$, $R_{25}$, $\sigma_{cg}$, and $\alpha=\sigma_z/\sigma_R$.  The accuracy of
the prediction depends on the details of the assumed mass distributions;
however, all four approaches we discuss exhibit systematic errors of roughly
35\% (68\% confidence interval).  Assuming our galaxy sample is representative
of the overall population \citepalias{2010ApJ...716..198B}, our calculations
demonstrate that one should {\em expect} an anti-correlation between $\Ssfre$
and $\qrw$, particularly for $\qrw^{1.5h_R} < 2$.

\subsection{ A Physical Link Between Disk Stability Level and Star Formation? }

Our use of empirical scaling relations to predict the anti-correlation between
$\Ssfre$ and $\qrw^{1.5h_R}$ suggests that this outcome is consistent with the
physical drivers of other morphological and dynamical outcomes of
late-type-galaxy evolution.  However, does the anti-correlation imply a physical
link between disk stability level and star formation?

The studies of \citet{2005ApJ...626..823L, 2006ApJ...639..879L} are particularly
relevant because the star-formation in their simulations is, in fact, driven by
gravitational instabilities, and our data are roughly consistent with their
predictions (Figures \ref{fig:sigsfr} and \ref{fig:sigsfg}).  However, most of
our galaxies are very stable, in the regime in which \citet{2005ApJ...626..823L}
find that it is difficult to form stars ($\qrwmin\gtrsim1.6$).  A comparison of
Figure 10 from \citet{2006ApJ...639..879L} with our Figure \ref{fig:sigsfr}
shows that our galaxies exhibit higher stability levels at the low star-forming
end.  Part of this discrepancy is due to systematic differences in our
calculation of the stability parameter (e.g., our applied corrections for disk
thickness); however, our data should yield relatively large stability values
even after removing these systematic differences.  Thus, just as seen via
ultraviolet radiation in the very extended parts of disks
\citep{2007ApJS..173..538T}, our measurements suggest that star-formation occurs
in locales with high stability levels.  One theory that addresses this
phenomenon is provided by \citetalias{2010ApJ...721..975O}.

\citetalias{2010ApJ...721..975O} \citep[see also][]{2011ApJ...743...25K} present
a model for star formation where the interstellar medium (ISM) is divided into
self-gravitating and diffuse components.  The pressure in the diffuse gas ---
regulated by heating, cooling, and supernova-driven turbulence --- is assumed to
be balanced by the vertical gravitational forces of the disk.  A fundamental
assumption is that the surface density of gravitationally bound clouds is
converted to stars over a star-formation timescale; however, this timescale
($\sim$2 Gyr) is much longer than the free-fall time.  The proportionality
$\Ssfr/\Sigma_g \propto \Sigma_\ast^{1/2}$ is a limiting behavior of their
model, assuming the star-formation rate in all galaxies is similarly affected by
chemical composition, turbulence, and magnetic fields.  We have shown in Figure
\ref{fig:sigsfg} that this proportionality is consistent with our data, albeit
with significant scatter.  We have not assessed the variations in metallicity,
turbulent pressure, or magnetic field strength in our sample, such that this may
explain some of the scatter seen in our data.

The correspondence of our data with the limiting behavior of the
\citetalias{2010ApJ...721..975O} prediction suggests that our galaxies form
stars at a rate that maintains the pressure balance in the diffuse ISM.  Such
star formation does not require large-scale gravitational instabilities, but it
does require the vertical gravitational field, largely effected by the stellar
component in most cases, to maintain the total pressure in the disk midplane.
Thus, star-formation can be active in the disks of our galaxies, despite our
rather large measurements of $\qrw$.  This may argue against a physical link
between $\qrw$ and $\Ssfre$ at large $\qrw$.

However, the consistency of the trend seen in Figure \ref{fig:sigsfr} with the
prediction of \citet{2006ApJ...639..879L} at low $\qrw$ is suggestive of the
role gravitational instabilities play in this small subset ($\sim$15\%) of our
sample.  Therefore, a direct physical link between the stability level of galaxy
disks and their star-formation activity may only be relevant to a small subset
of the galaxies in the local universe; however, the story is most likely very
different at earlier cosmic epochs \citep[see, e.g.,][]{2009MNRAS.397L..64A}.

\subsection{ Spiral Structure Effects }


A detailed analysis of the spiral-arm strength in our galaxy sample is beyond
the scope of this paper; however, Figure 9 from \citetalias{2010ApJ...716..198B}
shows that spiral arms are easily discernible in all of our sample galaxies.
From N-body simulations, we expect spiral structure to be more easily generated in
disks with lower $\mathcal{Q}_\ast$.  If we assume that the same is true for
$\qrwmin$, we might expect effects related to spiral arms to be more evident in
galaxies with lower $\qrwmin$.

In their appendix, \citetalias{2010ApJ...721..975O} discuss the effects of
spiral structure on their equilibrium model.  They find that the azimuthally
averaged (or disk-averaged) star-formation-rate surface density should be
significantly lower than the true value if the contrast between the arm and
inter-arm gas mass surface density is sufficiently high.  If this effect were
evident in our data, we may expect the residuals of our galaxy sample about
$\Ssfre \Sge^{-1}\Sse^{-1/2} = 10^{-3.25}$ $(G/{\rm pc})^{1/2}$ to be related to
the strength of the spiral structure and therefore to $\qrwmin$.  However,
Figure \ref{fig:sigsfgs} shows no such relation.  This may be because (1)
$\qrwmin$ is not a good proxy for the mass-loading of spiral-arm density waves,
(2) our galaxies have all been similarly affected by spiral structure such that
these properties have only changed the mean value of
$\Ssfre\Sge^{-1}\Sse^{-1/2}$, (3) the timescales involved in the passage or
lifetime of spiral arms is such that the equilibrium is never fully realized or
substantially altered within the spiral-arm regions, or (4) the scatter in
$\Ssfre \Sge^{-1}\Sse^{-1/2}$ caused by spiral arms or other physical processes
has obscured the relation.  Some of the scatter in our measurements of
$\Ssfre\Sge^{-1}\Sse^{-1/2}$ can be attributed to systematic error; however,
there is room for intrinsic scatter as well.  It is of great interest to
understand the scatter in $\Ssfre\Sge^{-1}\Sse^{-1/2}$, as it relates to spiral
structure and/or other physical properties that can affect how stars form.

\acknowledgements

We are very grateful to Eve Ostriker, Alessandro Romeo, and Marco Spaans for
providing comments on the submitted version of this document, which were very
helpful to our interpretation of the data.  We also thank the referee for a
careful reading of our paper, and for providing many insightful and helpful
comments.  Support for this work was provided by the National Science Foundation
(OISE-0754437 and AST-1009491), the Netherlands Organisation for Scientific
Research (614.000.807), NASA/{\it Spitzer} grant GO-30894, the Netherlands
Research School for Astronomy (NOVA), and the Leids Kerkhoven-Bosscha Fonds.

\appendix

\section{Dynamical Model}
\label{app:dynmodel}

Our dynamical model assumes a simple planar geometry of the disk defined in the
cylindrical coordinate system, $\{R,\theta,z\}$, with the disk inclination
defined as the angle $i$ between the disk normal and the line-of-sight (LOS), as
in \citet{2013ApJ...768...41A}.

We parameterize the {\em projected} stellar rotation curve using a PolyEx
function \citep{2002ApJ...571L.107G},
\begin{equation}
\pV_\ast = \pV_{\ast,0}\ \left[ 1 - \exp\left(\frac{-R}{h_v}\right)\right]\
\left[1+\eta \frac{R}{h_v}\right],
\end{equation}
such that the LOS stellar velocity is $V_\ast=\pV_\ast \cos\theta$.  We
parameterize $\sigma_R$, $\sigma_{ig}$, and $\Sigma_g$ using a sum of
exponentials, as in
\begin{equation}
\sigma_R = \sum_{j=0}^{N_e-1} \sigma_{R,j,0} \exp(s_j R),
\end{equation}
where the ``order'' of the function ($N_e$) is either one or two.  The ``order''
of $\sigma_R$, $\sigma_{ig}$, and $\Sigma_g$ has been chosen for each galaxy
individually based on a visual inspection of the azimuthally averaged data.
When $N_e=1$, we force $s_0<0$ (via our prior).  When $N_e=2$, we force $s_1 <
s_0$ to prohibit degeneracy between the components.  When $N_e=2$ for
$\Sigma_g$, one of normalizations is allowed to be negative to allow for a deficit
of gas in the galaxy center.

The SVE is assumed to be aligned with the cylindrical coordinate system such
that the LOS stellar velocity dispersion is 
\begin{equation}
\sigma^2_\ast = \sigma^2_R \left[ (\sin^2\theta + \beta^2\cos^2\theta)\sin^2i +
\alpha^2\cos^2i \right]
\end{equation}
where we have defined $\alpha=\sigma_z/\sigma_R$ and
$\beta=\sigma_\theta/\sigma_R$.  We assume $\alpha$ is independent of $R$.  The
gaseous velocity dispersion ellipsoid is assumed to be isotropic such that there
are no equivalent projection effects.

We assume the stellar orbits follow the epicycle approximation such that
\begin{equation}
\beta^2 = \frac{1}{2}\left( \partial_R \ln \pV_\ast + 1 \right),
\end{equation}
where 
\begin{equation}
\partial_R \ln x \equiv \frac{R}{x} \frac{\partial x}{\partial R}.
\end{equation}
Assuming that (1) the disk has a constant scale height, such that $\Sigma_{\rm
dyn} \propto \rho \propto \sigma^2_z$ \citep{1988A&A...192..117V}, and that (2)
there is no change in the $Rz$ velocity cross-terms with height $z$ above the
disk, we derive a simplified asymmetric-drift equation and calculate the
projected circular speed following
\begin{equation}
\pV_c^2 = \pV_\ast^2 + \sigma_R^2\sin^2i\left[ \beta^2 - 1 - 4
\partial_R\ln\sigma_R \right].
\end{equation}
We then calculate the projected rotation speed of the gas,
\begin{equation}
\pV_g^2 = \pV_c^2 + \sigma^2_{ig}\sin^2i(2 \partial_R\ln\sigma_{ig} +
\partial_R\ln\Sigma_g),
\end{equation}
following \citet{2010ApJ...721..547D}; for this correction we assume the
logarithmic derivative of the cold-gas mass surface density is the same as for
the ionized-gas.  The LOS gas velocity is $V_g=\pV_g \cos\theta$.

\section{Convergence of the Generative Model}
\label{app:converge}

Here we describe our usage of the stretch-move sampler
\citep{2013PASP..125..306F} in combination with a parallel-tempering algorithm
to sample the posterior probability of the dynamical model for each galaxy,
following the discussion in Section \ref{sec:genmodel}.

The stretch-move sampler uses a set of ``walkers'' in parameter space that are
advanced in series by proposing new positions based on the posterior
probabilities of the other walkers.  We define a ``draw'' from this sampler as
advancing one walker once and a ``full step'' as advancing all walkers once.
The parallel-tempering algorithm runs multiple stretch-move samplers in
parallel, where each sampler is assigned a ``temperature'', $T$, within a
temperature ladder usually following a geometric sequence.  This temperature
alters the target probability density to $P_T({\bth} | \mathcal{D},
\mathcal{H}) \propto \mathcal{L}^{1/T} P({\bth} | \mathcal{H})$,
such that a sampler with $T=1$ samples the nominal posterior and one with
$T=\infty$ samples a posterior that is identical to the prior.  Adjacent
samplers within the temperature ladder trade walkers $m$ and $n$ (selected
randomly without replacement) with probability
\begin{equation}
{\rm min}\left\{ 1, \left(\frac{1}{T_j}-\frac{1}{T_{j-1}}\right)
(\mathcal{L}_{j-1,m} - \mathcal{L}_{j,n}) \right\},
\end{equation}
where $T_j < T_{j-1}$, after applying a full step to each sampler.

We converge to the generative model of the data for each galaxy using the
parallel-tempering algorithm with five samplers, each with 200 walkers and with
a temperature ladder separated by factors of two, $T=\{16,8,4,2,1\}$.  The
parameter-space coordinates of the walkers are initialized only for the $T=1$
sampler, with the walkers of other samplers in the temperature ladder
initialized to exactly the same coordinates.  To begin, walkers are distributed
according to rough estimates of the model parameters and a normal error
distribution.  The error is assumed to be rather large so that the walkers
occupy a large volume.  For convergence, we iteratively generate a number of
sample sets.  Before beginning a new iteration, we ``reinitialize'' the $T=1$
sampler by selecting the 200 highest posterior probability samples from among
the unique parameter-space coordinates; the walkers in the remaining samplers
are reinitialized to exactly the same parameter-space coordinates.

A single execution of the MCMC is done in two stages.  (1) We draw $10^4$
samples --- perform a full step of each sampler 50 times --- and calculate the
autocorrelation time, $t_{\rm AC}$, of all the parameters.  The autocorrelation
time indicates how many times one needs to draw a sample before obtaining an
independent sample of the target probability density.  We iteratively add $10^4$
samples until $t_{\rm AC}$ for these samples is of order unity for all the model
parameters.  We then reinitialize the samplers.  (2) We draw $10^5$ samples ---
perform a full step of each sampler 500 times --- and iterate, by increasing the
scale parameter of the stretch-move sampler \citep[see equation 10
from][]{2013PASP..125..306F}, until the acceptance fraction is in the range
0.2--0.5.  If the acceptance fraction is too low, the MCMC yields samples that
are not sufficiently independent; if it is too high, the samples follow a random
walk through the parameter space.  In both extremes, the MCMC has not performed
its primary function, which is to produce independent samples of the parameter
space drawn in proportion to the posterior.  An acceptance fraction of 0.2--0.5
is recommended by \citet{2013PASP..125..306F}, and we typically achieve this
without need for iteration.

We run through these two stages multiples times to ensure that the MCMC has
passed its ``burn-in'' phase.  This is the phase when the MCMC is effectively
searching for the maximum probability density, before it starts to sample the
parameter space in proportion to the target probability density.


\end{document}